\documentclass[prb,aps,twocolumn,floats,superscriptaddress]{revtex4}
\usepackage{amsmath}
\usepackage{graphicx}
\usepackage{epsfig}

\def \>{\rangle}
\def \<{\langle}

\def\be{\begin{equation}}
\def\ee{\end{equation}}
\newcommand \bea {\begin{eqnarray} }
\newcommand \eea {\end{eqnarray}}

\newcommand{\nn} {\nonumber}

\begin{document}

\author{Pankaj Mehta}
\affiliation{Center for Materials Theory, Rutgers University,
Piscataway, NJ 08854}
\author{ Natan Andrei}
\affiliation{Center for Materials Theory, Rutgers University,
Piscataway, NJ 08854}

\title{ Scattering approach to Impurity Thermodynamics}

\begin{abstract}
Recently the authors developed a scattering approach that allows for a
complete description of the steady-state physics of quantum-impurities
in and out of equilibrium. Quantum impurities are described using
scattering eigenstates defined {\it ab initio} on the open, infinite
line with asymptotic boundary conditions imposed by the leads. 
The scattering states on the open line are constructed for
integrable quantum-impurity models by means of a significant
generalization of the Bethe-Ansatz which we call the Scattering
Bethe-Ansatz (SBA). The purpose of the paper is to present in detail
the scattering approach to quantum-impurity models and the SBA and
show that they reproduce well-known thermodynamic
results for several widely studied models: the Resonance Level model,
Interacting Resonance Level  model and  the Kondo model. Though the SBA 
is  more complex than the traditional Thermodynamic Bethe Ansatz (TBA)  
when applied to thermodynamical questions,
the scattering approach (SBA) allows access to an array of new
questions that cannot be addressed otherwise, ranging from scattering
of electrons off magnetic impurities to nonequilibrium dynamics.
\end{abstract}

\maketitle

\section{Introduction}

Recent advances in nanotechnology have allowed extensive
experimental study of quantum impurity systems out of equilibrium
in controlled, tunable settings \cite{Gordon}.  The impurities are
typically realized as quantum dots, tiny islands of
two-dimensional electron gas attached to leads via tunnel
junctions.  The number of electrons on the dot can be controlled
using a gate voltage since the hopping of electrons is impeded by
a large charging energy $U$.  When there is an odd-number of
electrons on the dot the upper-most energy level contains only a
single, unpaired electron, which behaves effectively as an
Anderson or Kondo impurity coupled to the two (or more) leads
playing the role of electron baths. Applying a potential
difference between the leads results in a current flowing across
the dot. A wealth of new experimental data has been collected in
recent years on quantum-impurities out of equilibrium in this
setting including current vs voltage curves and nonequilibrium
density of states (DOS) on the quantum dots
\cite{Amasha}. Nonetheless, 
a comprehensive theoretical understanding of the physics
of these models is lacking.

Quantum impurity systems are also the simplest examples of
strongly correlated electron systems, wherein interactions between
electrons are strong enough to result in new collective behaviors
which require a new set of degrees of freedom for their
description- the Kondo effect being a canonical example
\cite{Hewson}.  The strongly-correlated behavior is typically
characterized by a low energy scale such as the Kondo temperature
below which strong correlation physics dominates and perturbative
descriptions break down. One of the most fascinating new frontiers
in strongly-correlated systems is the study of such systems in
out-of-equilibrium situations. Quantum impurities are an ideal
experimental and theoretical setting for exploring the interplay
between nonequilibrium- and strongly-correlated dynamics due to
the relative simplicity of these models and the wealth of
experimental data available.

New theoretical questions arise in this context. Do sufficiently
large voltages suppress strong-correlations and thus kill the
Kondo effect? Do new scales, such as the decoherence scale, arise?
Does voltage effectively behave as a temperature?  How should one handle
intrinsically non equilibrium phenomena such as nonequilibrium
particle and energy currents or entropy production. What is the
effect of strong correlation the entropy production?

 Currently, the most commonly used technique to treat
quantum-impurities out of equilibrium is Keldysh perturbation theory
\cite{Rammer}.  The perturbative methods, however, are applicable only
in the high voltage regime and break down precisely where strong
correlations become important. As such, they are unable to answer the
interesting questions proposed above.  A variety of non-perturbative
techniques have been developed in order to capture the strong
correlation physics of quantum impurity models, mainly in the context
equilibrium physics. These include renormalization group methods,
techniques from bosonization and conformal field theory, and exact
solutions using the Bethe-Ansatz \cite{Hewson}. Most of these methods are no longer
applicable when the influence of nonequilibrium dynamics is comparable
to the strong correlations in the problem. This highlights the need
for new theoretical approaches that can probe the interesting
non-perturbative regimes \cite{Rosch, Schiller,Hershfield, Konik, Han,Borda}.

Recently we have introduced such a non-perturbative framework that
allows the description of a steady state out-of-equilibrium quantum
impurity system in terms of a time-independent scattering
formulation \cite{mehta}. A steady state ensues \cite{doyon} when the system is
{\it open}. Open systems must be defined directly on the infinite line
to allow an  in-flow and
out-flow of electrons and energy from the system.
The infinite volume limit, which
needs to be taken {\it ab initio}, provides a dissipation
mechanism. Under these circumstances the non-equilibrium steady state
can be described by a scattering eigenstate of the full hamiltonian,
an eigenstate defined on the infinite line with its asymptotic
behavior specified at the incoming infinity. In most cases  the
asymptotic boundary conditions are determined by the electron leads
\cite{mehta}.

We have subsequently also introduced a method, the Scattering
Bethe Ansatz (SBA), to construct those scattering eigenstates on
the infinite line for the Kondo model and other integrable
impurity models.  The traditional Bethe Ansatz, on the other hand,
which has been extensively applied to these models, is defined
with periodic boundary condition with periodicity $L$ (with $L$
subsequently sent to infinity).  This approach is appropriate to
{\it closed} systems and allows an efficient calculation of the
thermodynamic properties of the systems. However, it does not give
access to their scattering properties, nor to the non-equilibrium
physics.

 The scattering approach can also be applied under equilibrium
 conditions, when all baths are held at the same chemical potential, or
 in the case when only one lead is present.  The purpose of this paper
 is to study the Scattering Approach under these simpler circumstances
 and confront it with the conventional approach which can also be
 applied here.  We will show that the SBA reproduces known
 thermodynamical results  for quantum-impurity models. Nonetheless, 
as mentioned above, the algebraic Bethe-Ansatz and its finite
 temperature counterpart the Thermodynamic Bethe Ansatz, prove
 technically easier when calculating thermodynamics of
 quantum-impurity models. The real advantage of the SBA is that it
 allows us to harness the power of integrability to explore new
 questions about electron S-matrices and T-matrices important
 for understanding quantum-mechanical coherence and dephasing due to
 magnetic impurities \cite{Kaminski, Pierre}.  And, in a context which
 we will not further explore in this paper, SBA allows us to
 understand nonequilibrium steady-states in these models.

The paper is organized as follows. We start with a formal introduction
to the scattering approach to quantum-impurity models and the
scattering Bethe-Ansatz (SBA). Subsequently, we demonstrate our ideas
on the Resonant Level model where the physics is particularly simple
since the Hamiltonian is quadratic. Finally, we use the SBA to
construct scattering states to reproduce well-known $T=0$ equilibrium
results for Interacting Resonant Level and Kondo models. We conclude
the paper with some conjectures about the Kondo model that
significantly simplify the calculation of some impurity properties.

\section{The Scattering Formalism}

The basic idea underlying the scattering formalism is the
observation that a quantum impurity can be viewed as a localized
dynamical scatterer off which electrons from the attached leads or
host metal scatter.  The scattering changes the internal state of
both  the impurity and host electrons and leads to the generation
of strong correlations.  The standard procedure for treating such
problems is to set-up an initial state with wavepackets that
represent the incoming particles in the far past and evolve this
initial state for a very long time with a time-evolution operator
$U(t, t_o)= T\exp(-i \int_{t_o}^t H(t') dt')$ of the appropriate
interacting field-theory; $H$ is the Hamiltonian that describes
the particles and dynamical scatterer - in this case the quantum
impurity.  As the impurity is local, the interaction switches off
far away from the impurity and we can define 'in' or 'out' states
by specifying the asymptotic behavior in the far past or the far
future.  Namely, the 'in' , respectively, 'out' states are
eigenstates of the total Hamiltonian, satisfy the boundary
conditions that they tend to plane waves representing free
incoming particles in the $t\to -\infty$ and $t\to \infty$ limits
respectively.  The cross-section for a particular process is then
obtained by calculating the overlap of the ``in'' state with an
appropriate ``out'' state.  A recent application of these ideas to
quantum impurities is given in \cite{Zarand, Mehta05}.  While
scattering states are designed to allow access to the scattering
properties of the system, they also allow calculation of the
thermodynamic properties.

 The Hamiltonian for a quantum impurity
attached to a bath of free electrons is of the form:
 \be H = H_0 + H_{\rm int}=
\sum_{\alpha, \vec{k}} \epsilon_k \psi_{\alpha \vec{k}}^\dagger
\psi_{\alpha \vec{k}} + H_{\rm int} \ee with $\epsilon_k$  the
full three-dimensional dispersion of the electrons and $\alpha$
denoting the internal degrees of freedom of the electrons. The
term $H_{\rm int}$ describes the impurity and its interaction with
the bath of electron. Examples include the Kondo interaction,
\[
H_{\rm int} = J\sum_{a \vec{k}} \psi_{a \vec{k}}^\dagger \; \; (\vec{\sigma})_{a a'}\sum_{a'\vec{k'}}
\psi_{a' \vec{k'}} \cdot \vec{S},
\]
with $\vec{S}$ a localized spin representing the impurity, and the
resonance level model (RLM),
\[
H_{\rm int} = t \sum_{\vec{k}} (\; \psi_{ \vec{k}}^\dagger d +h.c.) +
\epsilon_d d^\dagger d,
\]
describing a local level at energy $\epsilon_d$ which hybridizes
with the bath electrons.

Standard manipulations \cite{Affleck95b} allow us to rewrite the
Hamiltonian as chiral 1-d field theories. Since only the
combination $\sum_{a \vec{k}} \psi_{a \vec{k}}^\dagger$ enters
into the interaction we can rewrite the theory in terms of the
field
\[
 \psi_{\alpha \epsilon}^\dagger = \int d^3 k \; \delta(\epsilon_k - \epsilon)
 \psi_{a \vec{k}}^\dagger
\]
as ($D$ denotes the bandwidth, namely the cut-off)
\[
H_0 = \int_{-D}^D d \epsilon \; \epsilon \; \psi_{\alpha \epsilon}^\dagger
\psi_{\alpha \epsilon}
\]
while the interaction terms take the form, $ J \int  d\epsilon \;
\psi_{a \epsilon}^\dagger \; \; (\vec{\sigma})_{a a'} \int
d\epsilon' \psi_{a; \epsilon'} \cdot \vec{S}$ or $ t (\int
d\epsilon \; \psi_{ \epsilon}^\dagger d +h.c.) + \epsilon_d
d^\dagger d$ for the Kondo or the IRLM Model respectively.
 Finally introducing a chiral fermion field
\[
\psi_{\alpha }^\dagger(x)= \int_{-D}^{D} e^{i \epsilon x}
 \psi_{\alpha \epsilon}^\dagger  \nu(\epsilon)^{-1/2}\;
\]
the kinetic term becomes
\[
H_0= -i \int_{-D}^{D}dx \; \psi_{\alpha}^\dagger(x) \partial_x
\psi_{\alpha}(x)
\]
while the field enters locally into $H_{\rm int}$, in the form
$\psi_{\alpha}^\dagger(0)$. As we are interested in the physics on
energy scales much smaller than the cut-off $D$, we consider only
universal results obtained in the limit $D \to \infty$.

{\bf Open vs Closed Boundary Conditions: } The scattering approach to
quantum impurity problems, by its very nature, is defined in infinite
systems, {\it without} boundaries. Physically, this is equivalent to
requiring that once incoming electrons scatter off the impurity they
do not return and scatter off the impurity again. We refer to infinite
size systems with no boundaries as ``open systems''. The infinite size
of the electron bath assures that the host metal or lead is a good
thermal bath. Real life systems are not infinite but possess
boundaries; our treatment is valid as long as the return time for the
electrons is much smaller than the system size.  The infinite size of
the system implies that scattering states are no longer normalizable
and in particular, the Feynman-Hellman theorem no longer holds
\cite{Feynman}. This will be important in understanding the results of
later sections when we construct eigenstates for the IRLM and Kondo
models.

 In this open system framework the nature of the incoming particles
that scatter off the impurity is specified by asymptotic boundary
conditions.  The incoming particles, far from the impurity, are
eigenstates of the free-electrons Hamiltonian $H_0$ and any eigenstate of
$H_0$ is a possible boundary condition describing what the incoming
particles look like.  Two different boundary conditions are
of primary interest: (i) when the incoming particles are
a Fermi sea, typically representing the host metal and (ii) when the
incoming particles are a Fermi sea and an excited quasi-particle.
The former allows for us to calculate thermodynamical properties from
scattering while the latter allows us to compute, in principle, single
particle $S$ and $T$ matrices.
These boundary conditions are depicted in Figure {\ref{fig:bceq}.
\begin{figure}[t]
\includegraphics[width=0.8\columnwidth, clip]{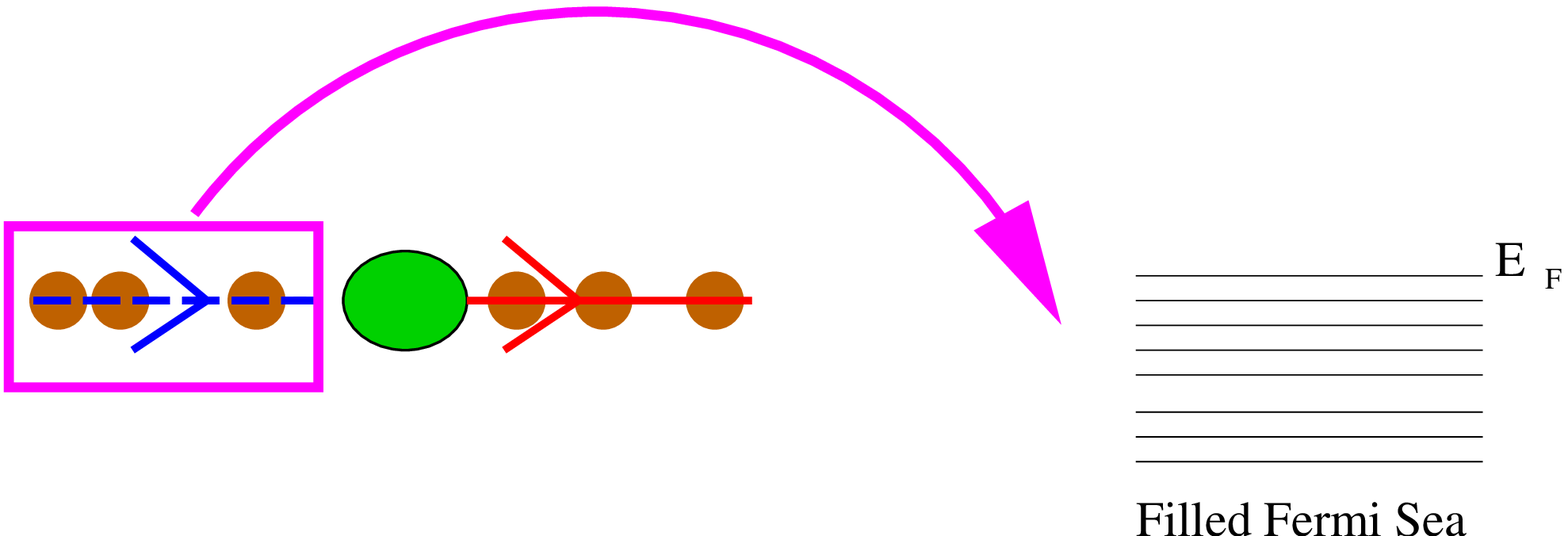}
\includegraphics[width=0.8\columnwidth, clip]{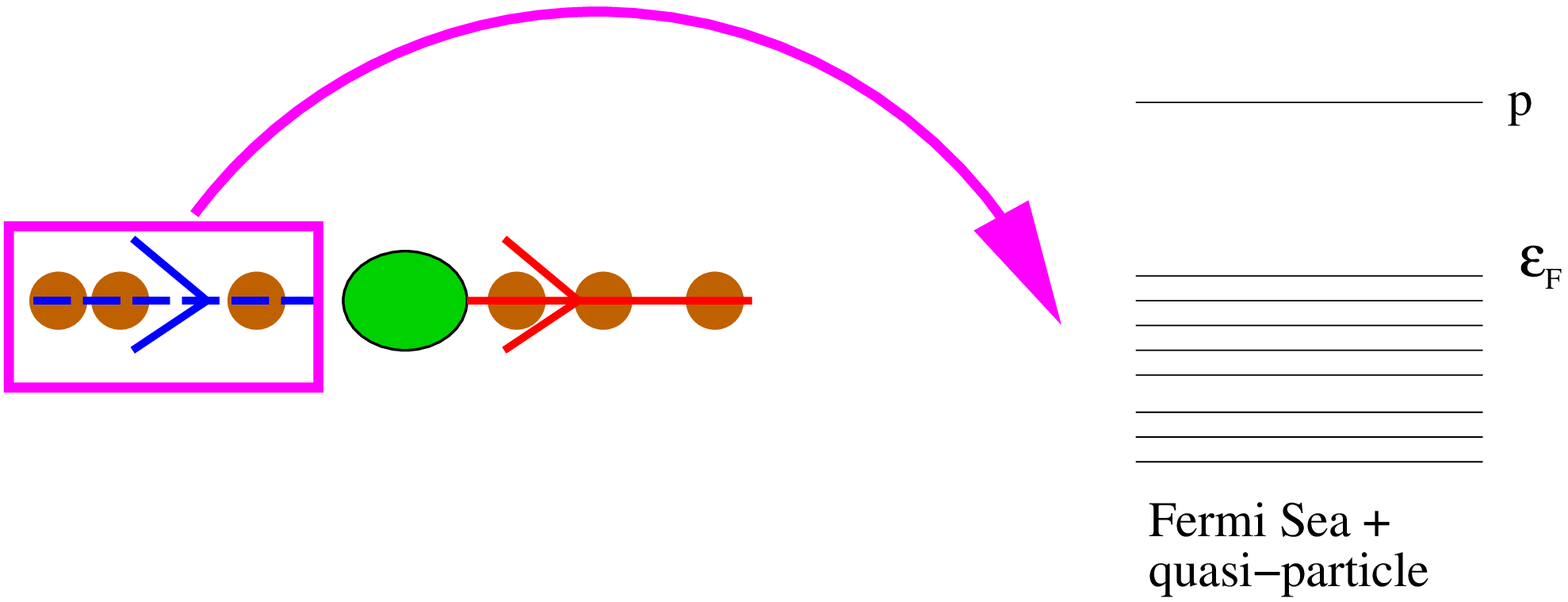}
\caption{
\label{fig:bceq}
The incoming particles in the chiral picture correspond to the
particles on the left of the impurity and the outgoing
scattered particles are  on the right. We take the convention that the
impurity is at $x=0$. Using this convention, incoming particles are
located on the negative x-axis, $x<0$, and outgoing particles to the
positive x-axis, $x>0$.  We typically consider two types of open
boundary conditions for incoming particles. In the first, the incoming
particles describe a filled Fermi sea. This is useful for calculating
thermodynamics. In the second, the incoming particles are a Fermi sea
and an excited quasi-particle. This allows us to calculate the
single-particle $S$ and $T$ matrices. }
\end{figure}

{\bf Time Dependent and Time Independent Formalisms: $T=0$:} There are
two different descriptions for scattering processes. In the
time-dependent description, the interaction between the conduction
electrons and quantum impurity is turned on in the far past, at $t=t_o$
and then adiabatically time-evolved using the Hamiltonian \be H=H_o +
e^{\eta t} H_{\rm int}\theta(t-t_o),
\label{chap1adH}
\ee
with $H_o$ describing the free electron bath and $H_{int}$ the
interactions between the quantum impurity and the incoming electrons,
i.e., between the dot and the leads.  Before $t=t_o$ the quantum
impurity is decoupled from the electron bath and the system is
described by an asymptotic boundary condition, an eigenstate of $H_0$
which we denote $|\Phi _o \rangle$ (at $T>0$ the system is described
by some density matrix $\rho_o$ describing decoupled leads and the
dot).  At later times, as the baths and the quantum-impurity evolve,
 the interaction is turned on adiabatically \endnote{
There are commonly two meanings of 'adiabaticity' in the
literature. In the first, adiabaticity means that when computing
physical quantities the limit $\eta \rightarrow 0$ is well
defined. This is equivalent to saying that one can actually turn on
the interaction infinitely slowly, with the limit $\eta \rightarrow 0$ taken
after the $L \to \infty$ limit.  This is the definition we use
here. The second, stronger meaning associated with adiabatic turning
on, is that the interaction is turned on so slowly that there are no
level crossings in the spectrum and there exists a one-to-one mapping
between the spectrum $H_0$ and $H$, now the $\eta \to 0$ limit is
taken before the infinite volume limit. We do not use adiabaticity in
this stronger sense. }  from the state $|\Phi _o \rangle$ under the
action of the time evolution operator
\bea
|\Psi(t) \rangle &=& U(t,
t_o)|\Phi _o \rangle \nonumber \\
 U(t, t_o) &=&T \{ \exp{(-i
\int_{t_o}^{t} dt^\prime H(t^\prime))}\}
\label{chap1defU}
\eea

We now wish to establish that a steady-state ensues after
sufficiently long time- long enough that all transients die out.
For this purpose one must show that the limit $ t_o \to -\infty$
exists, free of infra-red divergences.  This has been shown to be
the case for the Kondo model \cite{doyon} under the condition that
the infinite volume limit is taken first, i.e. the system is open
and the impurity is coupled to good thermal baths. The "openness"
of the system provided the dissipation mechanism necessary for the
steady state, allowing the high-energy electrons to relax and escape
to infinity.  The adiabatic limit $\eta \to 0$ is taken last,
allowing the smearing of the bath levels (level separation $\delta
\sim 1/L$ to take place turning the poles in the Green's function
into a cut).

 Under these circumstances $|\Psi (t)
\rangle$ become time-independent and describes a time independent
eigenstate which we denote $|\Psi \>_s$. Thus, we can also describe our state
in a time-dependent picture.
In the time-independent picture time is traded for
space and - for the chiral unfolded picture- the far past
corresponds to incoming particles located at distances
$x \ll 0$ and the far future to outgoing particles located at $x\gg 0$.
(see Figure {\ref{fig:bceq}). Under  both equilibrium and non-equilibrium
conditions (e.g. coupling to baths at different chemical potentials),
  the expectation value of any
operator $\hat{O}$ \be \< \hat{O} \> = \frac{\langle
\Psi|\hat{O}|\Psi\rangle_s} {\langle \Psi |\Psi \rangle_s}
\label{chap1timeindepexp}
\ee
is time independent.

  A stronger conclusion can be deduced which is central to our
  construction:   the state $|\Psi(0) \rangle = |\Psi \rangle_s$
  becomes,
by the Gellman-Low theorem \cite{GellMann}, an
eigenstate of the full Hamiltonian $H$, specified by the
initial condition $|\Phi_o\>$ that describes the electrons
in the far past ($ x \ll 0$).
In other words , the state $|\Psi\rangle_s$  is a scattering eigenstate
 of the full hamiltonian
 $H= H_0 + H_{\rm int}$, satisfying the Lippman-Schwinger equation,
\be
|\Psi\rangle_s
= |\Phi _o \rangle+ \frac{1}{E-H_0\pm i\eta }H_{\rm int}|\Psi\rangle_s
\label{chap1LSE}
\ee
with $|\Phi_o \rangle $ - the incoming state playing the role of a
boundary condition imposed asymptotically. The scattering state
$|\Psi\rangle_s$ can be viewed as consisting of incoming particles
(commonly taken to be a bath of free electrons) described by $|\Phi_o
\rangle $, and of scattered outgoing particles given by the second
term in the above equation. Once again two elements are required to
fully determine the system: a hamiltonian, $H$, and a boundary
condition, $|\Phi_o \rangle$, which describes the incoming scattering
state far from the impurity. Note that previously, in the
time-dependent picture, $|\Phi_o \rangle$ played the role of an
initial condition rather than a boundary condition.  As the impurity
is short ranged the scattering state $|\Psi \rangle_s$ must reduce to
the eigenstate $|\Phi_o\rangle$ when all the particles are far to the
left of the impurity.  This gives a prescription for calculating
scattering eigenstates for quantum-impurity problems. We must
construct an eigenstate of the full Hamiltonian $H$ with the
requirement that when all the electrons are to the left of the
impurity the eigenstate reduces to a prescribed eigenstate of $H_0$ describing
the free decoupled two baths and the impurity.  It is worth
emphasizing that we never explicitly solve (\ref{chap1LSE}). Instead,
we directly construct eigenstates of the full-Hamiltonian with
$|\Psi \rangle_s$ that are of the form described above.

{\bf Scattering formalism at finite Temperatures:} The above
discussion can be generalized to finite temperatures. Once again there
are two equivalent frameworks for quantum-impurity problems, a
time-dependent and time-independent. In the former, we proceed as in
the zero temperature case. We consider the quantum impurity and the
baths to be decoupled in the far past, at $t=t_o \to -\infty$,
adiabatically turning on  the coupling. The Hamiltonian is again given by
(\ref{chap1adH}).  The change from zero temperature is that the system
is no longer described by a single eigenstate but must instead by
described by a density matrix. At $t =-\infty$, the quantum
impurity is decoupled from the electron bath and the system is
described by the density matrix
 \be
 \rho_{0}= \exp{(-\beta H_o)} {\rm
\hspace{0.5in}}\,\,\,\,\beta= T^{-1}.
 \ee
At later times, the system
is described by time evolving the density matrix $\rho_{0}$ with the
time evolution operator
 \bea
 \rho(t) = U^{\dagger}(t, -\infty) \,
\rho_{0} \, U(t, -\infty)
\label{chap1timeevolvefiniteT}
\eea
with $U(t,-\infty)$  being understood as the limit $U(t, t_o \to -\infty)$.
The expectation value of an operator $\hat{O}$ can be calculated in
the usual manner by
\be
\< \hat{O} \> = \frac{ {\rm Tr}(\rho(t)\hat{O}) }{{\rm
    Tr}\rho(t)}
\label{rhotimedepexp}
\ee

 Again a time independent description can be given. Now the
boundary-conditions for our evolved density matrix $\rho_s$ is
provided by $\rho_0$: to the left of the impurity, we know that the
scattering density matrix $\rho_s$ must reduce to $\rho_0$.
Further, the finite temperature analogue of our zero temperature
condition that our scattering state $|\Psi \rangle$ be an eigenstate
of $H$ is requirement that the density matrix $\rho_s$ commute with
the full Hamiltonian in the limit $\eta \rightarrow 0$.

Thus, at $T > 0$ we consider the incoming states, $\{|\phi, m \>\} $, the
complete set of eigenstates of $H_o$ with energies $E_m$, distributed
with the probability of each state given by the Boltzman weight,
$e^{-\beta E_m^0}$,
\be
 \rho_0= e^{-\beta H_o}= \sum_{m} e^{-\beta E_m^0 }
|\phi, m\> \< \phi, m|,
\label{chap1defrho0}
\ee
 Using (\ref{chap1timeevolvefiniteT}), the time-independent
density matrix $\rho_s$ is
\bea
\rho_s &=& U(0, -\infty) \rho_o U^\dagger(0, -\infty) \nonumber \\
&=& \sum_{m}
e^{-\beta E_m^0} U(0,\infty) |\phi, m\> \< \phi, m|U^\dagger(0,
-\infty) \nonumber \\
&=&  \sum_{m} e^{-\beta E_m^0} |\Psi, m\> \,\<\Psi, m|
\label{chap1scatdensity}
\eea
where we have used (\ref{chap1defrho0}) and in the second line we have
defined the scattering state $|\Psi, m \> = U(0, -\infty)|\Phi, m\>$
with incoming particles describe by $|\Phi, m\>$. The
steady state physics is captured by the operator $\rho_s$ which
describes an ensemble of scattering states weighted by the Boltzman
factors determined by the energy of the incoming electrons. This form
for $\rho_s$ is consistent with the requirement that $\rho_s$ commute
with the Hamiltonian and reduce to $\rho_0$ to the left of the
impurity.  We can calculate the expectation value of an operator as in
(\ref{rhotimedepexp}) \be \< \hat{O} \> = \frac{ {\rm
Tr}(\rho_s\hat{O}) }{{\rm Tr}\rho_s}
\label{rhotimeindepexp}
\ee


\section{The Scattering Bethe-Ansatz}

We have shown in the previous section that the thermodynamic
properties can be obtained from scattering eigenstates defined
directly on the infinite line with {\it incoming} boundary conditions
imposed by the lead. In general,
constructing such eigenstates  is a formidable task due to the
strong correlations between particles, and is only carried out approximately.
But for a special class of models,
many of which have important direct physical application, the
many-particle eigenstates can be explicitly constructed using the
Bethe-Ansatz wavefunction form.

 The Bethe-Ansatz approach has a long history stretching back to
 Bethe's study of the Heisenberg model \cite{Bethe}.  The approach has been
 typically implemented on systems defined with periodic boundary
 conditions with respect to some finite length $L$. Subsequently the
 thermodynamic limit is achieved sending $L$ to infinity maintaining a
 constant density. If a field theory limit is to be taken, then a
 further scaling (or universality) limit is required.  By means of
 this ``Traditional Bethe Ansatz'' (TBA) approach the thermodynamics
 of several impurity models was studied in great detail, \cite{
 natanlectures, Affleck95b, TW}.

 Scattering, on the other hand, must by definition take place in
infinite systems with no boundaries - open systems in our
terminology. To compute scattering properties a different
formulation is required. This can be seen from several points of
view. To begin with, particles must come in from asymptotic
regions and after scattering occurs, escape again. Thus the system
must be open to allow the flow of particles and energy in and out
of the system. Furthermore, there must be a way to distinguish
between the incoming particles, typically {\it bare} particles,
eigenstates of $H_o$ but not of $H$, and the renormalized
quasiparticles that are the eigenstates of the latter but not of
the former.  Expressing the bare particles in terms of the
renormalized quasiparticles and vice versa lies at the heart of
the scattering theory.

Thus the traditional (or closed) Bethe-Ansatz (TBA) and the Scattering (or
open) Bethe-Ansatz (SBA) naturally address different sets of questions. The
natural questions that can be addressed using the first are
thermodynamical.  The TBA, by using a periodic system and requiring
wave-functions to be self-consistent, reproduces the full renormalized
excitation spectrum of a quantum-impurity model. With the knowledge of
the spectrum, one can use  statistical mechanics arguments to
calculate the thermodynamic quantities.  Boundary conditions (periodic
or otherwise) imposed on a finite length system are essential to this
approach. But all knowledge of the bare theory is lost.  As such, the
TBA is unable to tackle questions  about the scattering
properties of the quantum impurities. Scattering relies on working in
systems with bare particles and open boundary conditions- namely systems of
infinite extent with no boundaries.

There is a price to pay for working in open systems. The wavefunctions
are no longer normalizable and one does not have recourse to
thermodynamic concepts such as  free energy. Thus, while the SBA is
essential for analyzing scattering properties of quantum impurity
models, it is more difficult to extract the thermodynamics using
it. Table \ref{tab:ABASBA} summarizes the comparison between the two
approaches.
\begin{table}[t]
\centering
\begin{tabular}{||r|| r|r||}
\hline
\hline
& SBA & TBA \\
\hline
\hline
System & Infinite & Finite \\
\hline
Boundary condition & asymptotic (open) & periodic \\
\hline
Wavefunctions & used explicitly & not used \\
\hline
Thermodynamics & difficult & easy \\
\hline
Scattering   Properties  & possible & not possible \\
\hline
Nonequilibrium Generalization & Yes & No\\
\hline
\hline
\end{tabular}
\caption{Summary of differences between the Scattering Bethe-Ansatz (SBA)
and Algebraic Bethe Ansatz (ABA).}
\label{tab:ABASBA}
\end{table}

\subsection{The Bethe-Ansatz Wavefunction}

The central objective of the SBA is to construct on the infinite line
eigenstates of the Hamiltonian
\be H = H_0 + H_{\rm int}= -i\int_{-\infty}^{\infty}dx
\psi_{\alpha}^\dagger(x) \partial_x \psi_{\alpha}(x) + H_{\rm int}
\label{chap1genericHam2} \\
\ee
with the  condition that far away from the quantum-impurity
the incoming sector of the eigenstate reduces to a prescribed
eigenstate, $|\Phi\>$, of the free-electron Hamiltonian $H_0$. As such,
any scattering state must have a well-defined sense of incoming and
outgoing particles, with the incoming
electrons being to the left of the impurity ($x<0$) and the
outgoing scattered electrons those to its right.
 The state $|\Phi\>$ can be any eigenstate of $H_0$. We focus
in this paper on the case where $|\Phi\>$ is a Fermi-sea of
incoming particles. However, many other choices are possible. In
particular, to calculate S-matrices and T-matrices of
quantum-impurity Hamiltonian one can choose $|\Phi\>$ to be a
Fermi-sea with one incoming particle above the Fermi-sea (see
Figure \ref{fig:bceq}).

 The choice of $|\Phi\>$ describing incoming particles imposes an
asymptotic boundary condition on the full scattering eigenstate. In
general, imposing boundary conditions on our scattering states is
quite difficult. However, when the incoming particles are a free Fermi
sea, imposing the boundary-condition simplifies greatly.  The key to
this simplification is the observation that the natural basis for
Bethe-Ansatz wavefunctions is not the Fock basis, but a new ``Bethe
basis'' described extensively below.

The SBA constructs eigenstates of the Hamiltonian using wave-functions
of Bethe-Ansatz type \cite{Bethe}. The Bethe-Ansatz utilizes the
integrability of the Hamiltonian $H$ to divide multi-particle
scattering events into two-particle scattering events characterized by
the two-particle $S$-matrices, $S^{ij}$ derived from $H$. The
integrability of the Hamiltonian translates in this language into a
self-consistency condition on the two-particle S-matrices known as the
Yang-Baxter Equation (YBE) \cite{Yang}  ensuring that all multi-particle
interactions can be {\it consistently} broken-up into pair-wise
interactions. The consistent wavefunctions of the Bethe form, which we
collectively refer to as Bethe-Ansatz wavefunctions, are eigenstates
of the Hamiltonian.

We restrict our analysis to quantum-impurities coupled to
non-interacting electrons.  We further assume that particle number
is conserved, the Bethe-Ansatz wavefunctions all have a definite
number of particles, $N$, and there are only local interactions:
two particles can interact only if they are at the same point. To
write a Bethe-Ansatz wavefunction, it is necessary to divide the
configuration space into $N!$ regions according to the ordering of
the particles on the infinite line. For example, we can consider a
region where particle $5$ is to the left of particle $7$ which is
to the left particle $9$ etc., $(x_5 < x_7 < x_9 \ldots)$. Each
such region is labelled by a permutation $Q$ in the symmetric
group, $S_{N+1}$. Since a particle $i$ and $j$ can only interact
when they occupy the same position $x_i=x_j$, there are no
interactions in the interior of these regions. Within each region,
the Hamiltonian $H$ reduces to $H_0$ and the eigenfunctions are
sums of plane waves. The most general wave-function of the above
form is (with $x_0=0$ the position of the impurity)
\bea
|BA,\{p\}\> &=&\int dx_1 \ldots dx_N \,e^{i
\sum_{j} p_j x_j} \ \\ &&\sum_Q A^Q_{\alpha_1 \ldots
\alpha_N, \alpha_0}\theta(x_Q) \prod_{j=1}^N \psi_{\alpha_j}^\dagger(x_j)
|0, \alpha_0\>
\nn
\label{BAmomentawf}
\eea
where $\theta(x_Q)= \theta(x_{Q(1)} < x_{Q(2)} \ldots x_{Q(N)}<
x_{Q(0)})$ and $Q$ runs over all $N+1!$ permutations. The state $|0,
\alpha_0\>$ denotes the {\it drained} Fermi sea ($
\psi_{\alpha_j}(x_j)|0 \> =0$) and the state of the impurity.

When a boundary between two regions is crossed, two particles interact
(multi-particle interactions forbidden by Fermi statistics) and hence
the amplitude in the regions across the boundary are related by a {\it
two particle} S-matrix determined by solving the two-particle
Schrodinger Equation for the relevant Hamiltonian. The amplitude in a
region $Q$, $A_{\alpha_1\ldots \alpha_N}(Q)$, is related to the
amplitude in an adjacent region, $Q^\prime$, differing from it by the
exchange of neighboring particles $i$ and $j$, via the S-matrix
$S^{ij}$,
\bea
 A_{\alpha_1\ldots \alpha_N}^{Q^\prime} = (S^{ij})_{\alpha_1\ldots
\alpha_N}^{\beta_1 \ldots \beta_N} A_{\beta_1\ldots \beta_N}^{Q} =
(S^{ij})_{\alpha_i \alpha_j}^{\beta_i \beta_j} A_{\beta_1\ldots
\beta_N}^{Q}
\eea
 where in the second equality we have used the fact the two-particle
S-matrix $S^{ij}$ acts non-trivially only on the sectors of the
Hilbert space corresponding to particles $i$ and $j$.  In general, the
matrix relating the amplitude in the region $Q=I$, defined by $(x_1 <
x_2 <\ldots <x_N<x_0)$ is related to the amplitude in region $Q$,
$(x_{Q(1)} < x_{Q(2)} <\ldots <x_{Q(N)}<x_{Q(0)})$, by an S-matrix
$S^Q$ given by a product of two-particle exchange S-matrices $S^{ij}$
along the path leading from $I$ to $Q$. Since many paths can lead from
$I$ to $Q$ consistency requires that $S^Q$ be uniquely defined in a
path independent way. This is assured by the Yang-Baxter condition
\cite{natanlectures}. Thus, the Bethe-Ansatz
wavefunction can be written in terms of a single amplitude $A=A^I$ in the
region $Q=I$ and the S-matrices $S^Q$,
\bea |BA,\{p\}\> &=& \int
dx_1 \ldots dx_N \,e^{\sum_{j} p_j x_j} \label{chap1genericBAwf} \\
&&\sum_Q (S^Q A)_{\alpha_1 \ldots \alpha_N}\theta(x_Q) \prod_{j=1}^N
\psi_{\alpha_j}^\dagger(x_j) |0\> \nn.
 \eea

The energy of a
Bethe-Ansatz wavefunction (\ref{chap1genericBAwf}) is given
by $E=\sum_{j}p_j$. Note, however, that the Bethe-Ansatz wavefunction with
Bethe-Ansatz momenta, $\{ p_j \}$ is degenerate in energy with
all other Bethe-Ansatz wavefunctions $\{ p_j^\prime \}$ with
$\sum_j p_j^\prime = E=\sum_j p_j$. Thus, there are an infinite number
of degenerate Bethe-Ansatz wavefunctions of the same energy for any
Hamiltonian. Generically, a scattering state with energy $E$ is  a sum
of Bethe-Ansatz wavefunctions (\ref{BAmomentawf})
\be
|\Psi\> = \sum_{\{p \}; \sum_j p_j =E} C_{\{p \} }|BA,\{p \}\>,
\label{generalBAscatstate}
\ee
with  $C_{\{p\} }$ the amplitude in the scattering
state of the Bethe-Ansatz wavefunction $|BA,\{p\}\>$ and
the sum running over all sets of Bethe-Ansatz momenta $\{ p \}$ with
energy $E$.

\subsection{The Bethe-Ansatz Basis}

To construct scattering eigenstates for integrable quantum models
the Bethe-Ansatz wavefunction exploits the large degeneracy of the
{\it linearized} free electron gas. As taught in standard chapters
on degenerate perturbation theory the correct basis of states in a
degenerate subspace to perturb from is the one that diagonalizes
the perturbation, or equivalently, the one to which the  system
returns once the perturbation is turned off.  This is precisely
the intuition behind the ``Bethe basis'' of  a non interacting
field theory. A Bethe basis for a free electron gas is the basis
inherited from the interacting quantum-impurity theory when the
impurity is removed, or when the system is studied far from the
short range impurity. The basis is defined by the presence of a
non-trivial two particle S-matrix $S^{ij}$ between the right
moving free electrons in $H_0$. Indeed, a moment's reflection
shows that as the particles move with the same velocity (to the
right with $v_F=1)$ an S-matrix does not indicate interaction but
a choice of basis.

We now discuss the Bethe basis in more detail. For a
quantum-impurity model, there are two kinds of two-particle
S-matrices: those that describe electron-electron scattering,
which we denote $S^{ij}$, and those that describe
impurity-electron scattering which we denote $S^{0j}$. The
S-matrices $S^{ij}$ and $S^{0j}$ are determined by the impurity
interaction term $H_{\rm int}$ in (\ref{chap1genericHam2}) and the
Yang-Baxter consistency conditions.

 Imagine turning off the coupling to the
impurity in (\ref{chap1genericHam2}) so that $H_{int}=0$.
Then (\ref{chap1genericHam2}) reduces to the free-field
Hamiltonian $H_0$ and the electron-impurity S-matrix, $S^{0j}$ reduces
to the identity, $S^{0j} \rightarrow 1$. The electron-electron
S-matrix $S^{ij}$, however,  does not change.
This leads to the somewhat surprising conclusion that
Bethe-Ansatz wavefunctions of the form (\ref{chap1genericBAwf}) with
$S^{ij} \neq 1$ are eigenstates of the
free field Hamiltonian $H_0$.

This can be understood as follows.
Consider the first quantized version of $H_0$. In the two-particle
sector, the first quantized $H_0$ is given by $H_0=-i(\partial_{x_1}
+ \partial_{x_2})$. Notice that any wavefunction  of the form
\bea
|2;p_1, p_2;q\> &=& \int dx A_{\alpha_1\alpha_2}^q e^{i(p_1+q) x_1 +(p_2+q) x_2}
\nn \\
&&\psi_{\alpha_1}^\dagger (x_1) \psi_{\alpha_2}^\dagger (x_2) |0\>
\label{2partF}
\eea
is an eigenfunction of $H_0$ with energy $E=p_1+p_2$
(the $\{\alpha_i \}$ label the
internal degrees of freedom of the free electrons). Since $q$ can take
on any value, there is an infinite number of such states.
Any sum of eigenfunctions of the above form is also an eigenfunction
of the $H_0$ with energy $E$,
\bea
&&|2;p_1, p_2\> =  \\
&&\hspace*{-.2in}\sum_q \int dx_1 dx_2 e^{ip_1x_1 +i p_2 x_2} A_{\alpha_1 \alpha_2}^q
e^{iq(x_1-x_2)} \psi_{\alpha_1}^\dagger (x_1) \psi_{\alpha_2}^\dagger (x_2) |0\>\nonumber \\
&=& \int dx_1 dx_2 e^{ip_1x_1 + ip_2 x_2}f_{\alpha_1\alpha_2}(x_1-x_2)\psi_{\alpha_1}^\dagger (x_1) \psi_{\alpha_2}^\dagger (x_2) |0\> \nn
\label{2partBA}
\eea
where to go from the first line to the second line
we have used the fact that $ \sum_{q} A_{\alpha_1 \alpha_2}^q
e^{iq(x_1-x_2)}$ is the general expression for the
Fourier transform of an arbitrary function, $f(x_1-x_2)$, of $x_1-x_2$.
Thus, due to the large symmetry of the free electron problem,
there is an infinite number of degenerate two-particle eigenstates for
$H_0$. The above argument easily generalizes to more than two
particles: any function of the form
\bea
|N\>= \int dx_1 \ldots dx_N e^{\sum_{s=1}^N p_s x_s}
\prod_{i<j}f_{\alpha_i\alpha_j}(x_i-x_j) \nn \\
\prod_j \psi_{\alpha_j}^\dagger (x_j) |0\>
\label{Npartdiff}
\eea
is an $N$-particle eigenstate of $H_0$. Since
$\theta(x_Q)= \prod_{i<j}\theta(x_{Q(i)-Q(j)})$, is of
that form \ref{Npartdiff}
we conclude that the most general $N$-particle
Bethe-Ansatz wavefunction with $S^Q$ a product
of electron-electron S-matrices, $S^{ij}$,
\bea
|N , BA\> = \int d\vec{x} e^{i\sum_j p_jx_j}(S^Q
A)_{\alpha_1 \ldots \alpha_N} \theta(x_Q) \nn \\
\psi_{\alpha_1}^\dagger(x_1)\ldots \psi_{\alpha_N}^\dagger(x_N).
\label{NpartBAwf}
\eea
is an eigenstates of $H_0$.  However, for $S^{Q} \neq 1$ (which
implies $S^{ij} \neq 1$), it is clearly not of the usual Fock-basis
form,
\be
|N,F\> =\int d\vec{x} e^{i\sum_j p_j x_j}A_{\alpha_1 \ldots \alpha_N}
\psi_{\alpha_1}^\dagger(x_1)\ldots \psi_{\alpha_N}^\dagger(x_N) |0 \>.
\label{NpartFockwf}
\ee
The different choices for $S^{ij}$, and in turn $S^Q$, correspond
to different 'Bethe-Ansatz' bases for free electrons.
The choice of $S^{ij}$ imposed by the
impurity interaction corresponds to working in a particular "Bethe-Ansatz"
basis for the problem. The usual Fock basis corresponds to the choice
$S^{ij}=1$.

We now proceed to discuss the relationship between the Bethe basis,
with $S^{ij} \neq 1$ and the Fock basis $S^{ij}=1$. We denote, for a
particular choice of a consistent set of matrices $S^{ij}$, the
resulting Bethe-Ansatz wavefunctions by $\{ |BA \>_n\}$ where $n$
enumerates all possible choices for the $\{ p \}$ and $A_{\alpha_1
\ldots \alpha_N}$ in (\ref{NpartBAwf}).  The set of
Bethe-Ansatz wavefunctions $\{ |BA \>_n\}$ form a complete basis for
our Hilbert space of $H_0$ in the limit of infinite size and particle
number. In quantum mechanics, different basis for the Hilbert space
are related by unitary transformation. Thus, we can formally define an
operator $U$ that relates the Fock basis $\{ |F\>_m \} $ to the BA
basis $\{ |BA \>_n \}$. $U$ maps states in the Fock basis
(\ref{NpartFockwf}) to states in the Bethe-Ansatz basis
(\ref{NpartBAwf}). In general, the matrix $U$ relating the two basis
is quite complicated since a single state in the Fock basis $|F\>_i$
maps onto a sum of wavefunctions of the Bethe-Ansatz form $|F\>_n
\rightarrow \sum_{m} U_{nm} |BA\>_m$. For example, in (\ref{2partBA})
we saw that the two particle eigenstate is actually a sum over many Fock states
of type (\ref{2partF}).

However, $U$ simplifies greatly if we restrict ourselves to asking how
the {\it ground state of $H_0$} in the Bethe-Ansatz and Fock basis are
related.  For a systems with unique ground-states, $U$ must map the
Fock basis ground state, $|N,F\>_{gs}$ to the ground state in the
Bethe-Ansatz basis $|N,BA\>_{gs}$.  Thus, the ground state in the Fock
basis maps to a single wavefunction of the Bethe-Ansatz form
(\ref{NpartBAwf}). Since the ground state of $H_0$ is a free
Fermi-sea, it follows that a Fermi-sea can be represented by a single
Bethe-Ansatz wavefunction. In the sections that follow, we will
restrict ourselves to this case where we represent a free Fermi sea,
the ground state of $H_o$ in both basis.

\subsection{Imposing Asymptotic Boundary-Conditions}

The goal of the SBA is to construct eigenstates of the Hamiltonian
(\ref{chap1genericHam2}) satisfying the asymptotic boundary-condition
that the incoming particles are a prescribed eigenstate, $|\Phi\>$, of
$H_0$. We focus on the simplest case when incoming particles come from
a bath and are a free Fermi-sea. Central to the imposition of any
boundary-condition on the fully interacting Bethe-Ansatz wavefunctions
is the observation that these wave functions pick a particular
Bethe-Ansatz basis for the free Hamiltonian $H_0$. Thus, the boundary
condition, typically formulated in the Fock basis, must be
reformulated in the natural basis for the scattering state
wavefunctions, the Bethe-Ansatz basis. The antagonism between the Fock
basis, natural for boundary-conditions, and the Bethe-Ansatz basis,
natural for wavefunction is at the heart of many of the SBA. We
discuss only the zero-temperature case. The generalization to finite
temperatures is straightforward.

Recall that the incoming electrons in our chiral picture are
electrons to the left of the impurity, $x<0$ (see Figure
\ref{bceq}). Thus, the asymptotic boundary condition requires that
the scattering state reduce to the eigenstate of $H_0$, $|\Psi\>
\rightarrow |\Phi_o \> = |\Phi\>_{baths} \otimes |\alpha_d \>$, when
all particles are to the left of the impurity, $\{ x_j \} <0$, with
$|\Phi \>_{bath}$ a state describing a Fermi sea of free electrons. In
general, the scattering state $|\Psi\>$ is a sum of wavefunctions of
the Bethe-Ansatz form (\ref{generalBAscatstate}). The amplitudes of
the different Bethe-Ansatz wavefunctions $C_{\{p_j\}}$ are determined
by the asymptotic boundary condition.  It was argued in the last
section that the $|\Phi \>_{baths}$ can be written using a {\it
single} Bethe-Ansatz wavefunction of the form (\ref{NpartBAwf}). Thus,
in the case where the incoming particles are described by
$|\Phi\>_{baths}$, our scattering state $|\Psi\>$ can also be
described by a single Bethe-Ansatz wavefunction.  The incoming
electron corresponds to the regions in the wavefunctions of the form
$\theta(x_{Q^\prime};x_0) \equiv \theta(x_{Q^\prime(1)}
<x_{Q^\prime(2)}<\ldots < x_{Q^\prime(N)}< x_0)$ with $Q^\prime$ a
permutation of the $N^e$ electrons in the problem. Since there are no
electron-impurity scattering events in these regions, $S^{Q^\prime}$
can be written entirely in terms of the electron-electron scattering
matrix $S^{ij}$ and the scattering state $|\Psi\>$ reduces to
$|\Psi^-\>$ when all electrons are to the left of the impurity,
 \bea
|\Psi \rangle &\to& |\Psi^-\>=\int dx_1 \ldots dx_N \,e^{i\sum_{j} k_j
x_j} \\ && \sum_Q^\prime S^{Q^\prime} A_{\alpha_1 \ldots \alpha_{N_e};
\alpha_0} \theta(x_{Q^\prime};x_0) \prod_{j=1}^N
\psi_{\alpha_j}^\dagger(x_j) |0\>.\nn
\eea
 The right hand side is
precisely of the form (\ref{NpartBAwf}). We therefore conclude that
$|\Psi \>$ reduces to eigenstate of $H_0$ in the Bethe-Ansatz basis
when all particles are to the left of the
impurity. This leads to the observation that when the incoming
particles are a free Fermi sea, {\it imposing the asymptotic boundary
conditions corresponds to choosing the amplitude $A_{\alpha_1 \ldots
\alpha_N}$ and the Bethe-Ansatz momenta $\{ p_j \}$ for a single
wavefunction of the form (\ref{NpartBAwf})} such that $|\Psi^-\>$
describes a Fermi sea.

 As is usual in the
Bethe-Ansatz, we do not  seek to determine the BA momenta $\{ p_j
\}$ in the thermodynamic limit, computing, instead,  the distribution function for the BA  momenta,
$\rho(p)$. For an infinite system, the distribution $\rho(p)$
and the amplitude $A_{\alpha_1 \ldots \alpha_N}$
are {\it independent} of the procedure used to arrive at them \cite{natanlectures}. This
observation allows us to find  $\rho(p)$ and $A_{\alpha_1 \ldots \alpha_N}$
using an auxiliary Algebraic Bethe Ansatz problem for a system of free
electrons on a finite ring of length $L^\prime$
with Hamiltonian $H_0$ and
two-particle S-matrices, $S_e^{ij}$. In the limit $L^\prime \rightarrow \infty$,
the distribution function for the
BA momenta and amplitude in the auxiliary problem will coincide with
those of the physical system.  $\rho(p)$ and $A_{\alpha_1 \ldots \alpha_N}$
are obtained in the auxiliary problem in the usual way by requiring that
the wavefunction be periodic. In particular, the amplitude
$A_{\alpha_1 \ldots \alpha_N}$ and the BA momenta $\{p_j\}$ must
satisfy the auxiliary Bethe-Ansatz equations,
\bea
e^{ip_jL^\prime}A_{\alpha_1 \ldots \alpha_N}= S^{jj-1}\ldots S^{j1}S^{jN}\ldots S^{jj+1}
A_{\alpha_1 \ldots \alpha_N} \nn
\eea
This program is carried out explicitly for the IRLM and Kondo models
in the appendix.

To summarize, the imposition of the asymptotic boundary condition on
the incoming particle greatly simplifies in the special case where the
incoming particles are a free Fermi-sea. The scattering state $|\psi
\>$ can be described by a single Bethe-Ansatz wavefunction and
the imposition of the boundary condition is reduced to finding the amplitude
 and BA momenta  for this BA wavefunction. These are found by
using the TBA to treat the auxiliary problem of free electrons on a
finite ring of length $L^\prime$
in the appropriate Bethe-Ansatz basis. The amplitude and the BA
momenta of the auxiliary problem coincide with those of the
scattering state in the limit $L^\prime \rightarrow \infty$.

\subsection{Computing with Scattering States}

Thus far, we have discussed the explicit construction of the
scattering states $|\Psi \>$ for integrable quantum impurity
models. We proceed now to compute the expectation values
of physical quantities in the 
scattering eigenstates using (\ref{chap1timeindepexp})
 \be
\<\hat{O}\> = \frac{\< \Psi|\hat{O}|\Psi\>} {\<\Psi|\Psi\>}.  
\ee 
The calculations of expectation values are greatly simplified because we
work directly with infinite systems. Technically, this is because for
strictly infinite systems, we can ignore all but one term in the
Slater-determinants occurring in the above expression. This
simplification is the mathematical expression of the
 physics for infinite systems: electrons that scatter off the
impurity ``leave'' the system and never return to scatter off the
impurity again.

 Consider first the overlap between two Bethe-Ansatz
wavefunctions. Since they are given as a sum of plane waves in each
region $Q$, the overlap of two such-wavefunctions, $\<
BA,\{p_j\}|BA,\{k_j\}\>$, is  (suppressing the internal index
$\alpha_j$ for notational brevity)
\bea
\sum_{Q,\tilde{Q}}\int
d\vec{x} d\vec{y}\,e^{i\sum_j(k_jx_j-p_jy_j)}
\theta(x_Q)\theta(y_{\tilde{Q}}) \nn \\A(Q)A(\tilde{Q})
\<0|\prod_{s=1}^N \psi(y_s) \prod_{j=1}^N \psi^\dagger(x_j) |0\>
\eea
The Fermions field give rise to a Slater determinant
 \bea
\sum_{Q,\tilde{Q}, S} (-1)^{sgn(S)} \int d\vec{x}
d\vec{y}\,e^{i\sum_j(k_jx_j-p_jy_j)} \theta(x_Q)\theta(y_{\tilde{Q}})
\nn \\ A(Q)A(\tilde{Q}) \prod_{j=1}^N \delta(x_{S(j)}-y_j)
 \eea
 where $S$ is a permutation of the $N$ particles.  Integrating over
$\vec{y}$, we have

\begin{align}
&\< BA,\{p_j\}|BA,\{k_j\}\> = \label{temp232} \\
=&\sum_{Q, S} (-1)^{sgn(S)}
\int d\vec{x} \,e^{i\sum_j(k_jx_j-p_jx_{S(j)})}
\theta(x_Q)A(Q)A(QS^{-1}) \nonumber \\
 =& \sum_{Q, S} (-1)^{sgn(S)}
\int d\vec{x} \,e^{i\sum_j(k_j-p_{S^{-1}(j)})x_j}
\theta(x_Q)A(Q)A(QS^{-1}) \nn
\end{align}

Thus, we see that this expression is the norm of plane waves
integrated over a region $\theta(x_Q)$. As is usual we regularize plane waves
by first placing the system in a box of size $L$ whose size is then
taken to infinity at the end of the calculation.
This allows us to consider the simpler
problem of plane-waves
\be
\lim_{L \rightarrow \infty} \int_{-L/2}^{L/2}
dx e^{i(k_j-p_j)x_j } \theta(x_1<x_2 \ldots< x_N).
\ee
It is straightforward to show that the leading order contribution
in $L$ to this integral is
$L^N/N!$ which occurs only if the two sets of Bethe-Ansatz momenta are identical
$\{k_j \}= \{ p_j \}$. This is the statement that plane waves
are 'orthogonal' even on a region $\theta(x_Q)$ for an infinite system.
Thus, for infinite size systems
we can ignore all terms in (\ref{temp232}) where the $k_j \neq
p_{S^{-1}(j)}$ for all $j$..
This leads to great technical simplifications as we only need to keep
terms in the sum (\ref{temp232}) where $Q=1$. Similar, simplifications
occur when computing the expectation value of an operator $\hat{O}$
between Bethe-Ansatz wavefunctions.

\section{Scattering Approach to the Resonant Level Model}

In this section, we will apply the scattering framework to a
quadratic model, the Resonance Level Model (RLM). Despite its
simplicity there is much interest in this model because 
it describes the strong
coupling physics of the Kondo model.  It will be shown that our results
agree with other approaches to this model such as Keldysh or Landauer
which can be carried out completely in this quadratic case. In the
next section we shall apply our approach to a fully interacting model
with strong correlations.

 The Hamiltonian for the RLM,
\bea H_{\rm RL}&=& H_0 + H_{RL {\rm int}}
\label{chap1RLM} \\ H_0\;&=&-i\int dx \, \psi^\dagger(x) \partial_x
\psi(x) \nonumber \\ H_{RL{\rm int}}&=&t(\psi^\dagger(0)d + h.c.) +
\epsilon_d d^\dagger d, \nn
 \eea
 describes a local level $d^\dagger$
onto which electrons can hop on and off. The energy of the level
(relative to the Fermi energy) is controlled by $\epsilon_d$, related
to the magnetic field in the anisotropic Kondo \cite{Imry}. Notice,
that we have already projected to one dimension and there are only
right moving chiral electrons. As explained in the last sections, in
the chiral picture, the free incoming electrons are located to the
left of the impurity, $x < 0$, and the scattered outgoing electrons
are to the right of the impurity, $x>0$.  The RLM serves as a good
pedagogical introduction to the scattering framework for quantum
impurity models since it is quadratic model and we will not have to
resort to the full machinery of the scattering Bethe-Ansatz.

\subsection{RLM at $T=0$: Thermodynamical Properties}

Consider first the zero temperature thermodynamics.
We must construct a
'in' scattering state, $|\Psi  \>_s$, describing incoming
electrons
from the host metal scattering off the impurity. The
scattering state $|\Psi \>_s$ is an eigenstate of the full
Hamiltonian (\ref{chap1RLM}) such that when all the particles are to the
left of the impurity $|\Psi \>_s$ reduces to an eigenstate
$|\Phi_o \>$ of $H_0$ describing a Fermi see (see Figure\ref{fig:bcfigeq}).
\begin{figure}[t]
\includegraphics[width=1.0\columnwidth, clip]{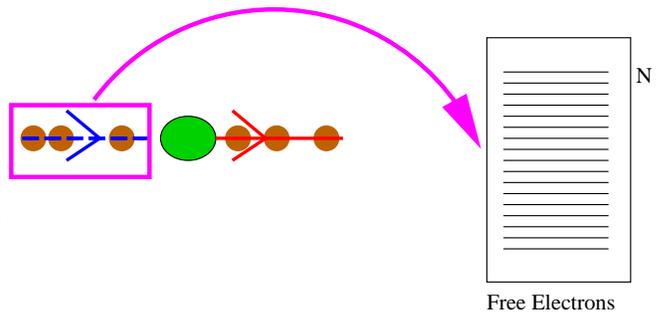}
\caption{
\label{fig:bcfigeq}
The scattering state $|\psi\>$ describes a quantum-impurity where
the incoming particles ($\{x_i <0\}$) are a free Fermi-sea with $N$ particles.  }
\end{figure}

The
RLM Hamiltonian (\ref{chap1RLM}) conserves total particle number. Hence, we can work in a
sector of the Hilbert space with a definite number of
particles, $N$.  Beginning with $N=1$, the most general single particle eigenstate is of the form
\be
|1p\>_s= \left(
\int_{-\infty}^{\infty} dx e^{ipx} g_p(x)\psi^\dagger(x) + e_pd^\dagger \right)|0\rangle
\label{chap1RLsinglegeneral}
\ee
Applying the Hamiltonian leads
to Schrodinger equation
\bea
\partial_x g_p(x) + Ve_p\delta(x) &=& pg_p(x)\\
tg_p(0)+\epsilon_d e_p &=& p e_p.
\label{chap1RMSE}
\eea
Taking the ansatz that $g_p(x)$ is of the form $g_p(x)=
A\theta(-x) + B\theta(x)$ and inserting this into the above equation,
one has, using the regularization scheme
$\delta(x)\theta(x)=\frac{1}{2} \delta(x)$ \cite{natanlectures},  that
\bea
\frac{B}{A} = \frac{1+ i\frac{t^2}{2(p-\epsilon_d)}}
{1-i\frac{t^2}{2(p-\epsilon_d)}} \equiv e^{i\delta_p}
\label{chap1defphaseshift}
\eea
Thus, the most general single particle eigenstate is  given by
\bea
|1p\>_s &=& \left(\int dx e^{ipx} A(\theta(-x) + e^{i\delta_p}\theta(x)) \nn
+e_p d^\dagger\right)|0\> \\
e_p &=& \frac{tg(0)}{(p-\epsilon_d)}= \frac{t(1+e^{i\delta_p})}{2(p-\epsilon_d)}
\label{chap1RL1partES}
\eea
where to get the second equation we have used (\ref{chap1RMSE}) and
the aforementioned regularization scheme. For future reference it
will be helpful to define the single particle scattering state creation operator
\be
\alpha_p^\dagger(x)=(\theta(-x) + e^{i\delta_p}\theta(x)) \psi^{\dagger}(x)
+\delta(x) e_p d^\dagger
\label{chap1defalpha}
\ee

Since the Hamiltonian (\ref{chap1RLM}) is quadratic, a $N$-particle
eigenstate is given by a tensor product of single
particle eigenstates. The most general $N$-particle eigenstate is
of the form
\be
|\Psi \> = \prod_{j=1}^{N}\otimes |1p_j \>= \prod_{j=1}^{N} \int d\vec{x} e^{i \sum_j p_j x_j} \alpha_{p_j}^\dagger(x_j)|0\>
\label{chap1RLstate}
\ee
Notice that we have not yet specified the momenta $\{p_j \}$ of the
state. Since we wish to construct a scattering eigenstate, these
momenta must be chosen to satisfy the boundary condition that when
all particles are to the left of the impurity our eigenstate reduces
to an eigenstate of $H_o$ describing the incoming electrons of the host metal
or lead at thermal equilibrium (see Figure \ref{fig:bcfigeq}).
At zero temperature, this means that the
scattering state must reduce to $|\Phi_o \> =|\Phi \>_{baths}  \otimes |\phi_d \>$,    when all
particles are to the left of the impurity.  Here  $|\phi_d \>$ describes some impurity state
and  $ |\Phi \>_{baths} $  describes a free Fermi sea
\be
|\Phi \>_{baths}= \int dx_1\ldots dx_N e^{i\sum_{j=1}^{N}p_j x_j}
\psi^\dagger(x_1)\ldots \psi^\dagger(x_N)|0\>
\label{chap1RLphi}
\ee
 under the {\it additional condition} that the momenta of
the particles $\{ p_j \}$ be distributed according to the Fermi-Dirac
distribution function. Since we are interested in the limit where the
number of particles goes to infinity, it is sufficient to specify the
distribution of the momenta instead of the individual values of the
momenta themselves.

The single particle eigenstate (\ref{chap1RL1partES}) consists
of an incoming particle,
$\int dx \theta(-x) e^{ipx}\psi^\dagger(x) |0\>$,
and an outgoing scattered wave, $\int dx \theta(x)
e^{i(px+\delta_p)}\psi^\dagger(x) |0\>$. Since the multi-particle scattering
state (\ref{chap1RLstate}) is a tensor product of the single particle
state, when all particles are to the left of the impurity $|\psi\>_s$
reduces to
\begin{align}
&|\Psi \>_s \to \\
&\int dx_1 \ldots dx_N
\prod_{s=1}^N\theta(-x_s)e^{i \sum_{j=1}^N p_j
x_j}\psi^\dagger(x_1)\ldots \psi^\dagger(x_N)|0\>.
\label{chap1psiphi}  \nn
\end{align}
If we choose the momenta $\{ p_j \}$ to be distributed according to
the Fermi-Dirac distribution, (\ref{chap1RLstate}) reduces to
the expression  for $|\Phi \>_{baths} $. Hence, our
scattering state is  given by (\ref{chap1RLstate}) with the
requirement that the momentum distribution
of the electrons be chosen according to the Fermi-Dirac distribution.

{\bf Expectation values}
We can calculate the expectation value of operators for the RL model
using (\ref{chap1timeindepexp}). We are interested in calculating the dot
occupation $n_d = \< d^\dagger d \>$.  Since the
multi-particle eigenstate (\ref{chap1RLstate}) is  a tensor product of single
particle eigenstates (\ref{chap1RL1partES}), it is useful to prove some
identities about single-particle scattering states. We regularize our system,
as is usual in scattering theory, by placing the system in a box of length $L$.
The physical system corresponds only to the $L=\infty$ limit and finite $L$
properties are {\it not} well defined.
A straightforward calculations yields
(without loss of generality setting $A=1$ in (\ref{chap1RL1partES}))
\bea
&&\<1k|1p\> =\label{chap1RL1partnorm} \\
 && L\delta_{pk}
+ \left[ |e^p|^2
\delta_{pk}+ \left(\frac{1-e^{i(\delta_p-\delta_k)L}}{i(p-k)}\right)
(1-\delta_{pk}) \right] \nn
\eea
and
\bea
\<1k|d^\dagger d|1p\> &=& e_k^*e_p
\label{chap1RL1partident}
\eea
Thus the overlap of states
with the same momenta is of higher order in $L$ than those with
different momenta, so that plane waves are an
orthogonal basis for infinite size systems. In the scattering framework which
works directly with infinite size systems, it is sufficient to consider
overlaps only of single-particle states with the same momenta.

Consider now the dot occupation. To leading order in $L$, one finds,
 combining (\ref{chap1RL1partident}), (\ref{chap1RL1partnorm}),
 (\ref{chap1RLstate}), and (\ref{chap1timeindepexp}), that the
 occupation is given by
\be
 \<n_d \>=\frac{1}{L}\sum_{j=1}^N
 \frac{|e_{p_j}|^2}{|1+e^{i\delta_{p_j}}|^2}
 =\frac{1}{L}\sum_{j=1}^{N_1} \frac{2\Delta}{\Delta^2 +
 (p_j-\epsilon_d)^2}
 \ee
 with $\Delta=t^2/2$, where to go from the
 first to the second line we have used the explicit forms of $e_p$ and
 $e^{i\delta_p}$.  Since, we are interested in the infinite size limit
 $N,L \rightarrow \infty$, we can replace the sum by an integral over
 the distribution of incoming electrons which is given by the
 Fermi-Dirac distribution function, $\theta(\epsilon_f - p)$ to yield
\be
 \<n_d\>= \int dp \, \theta (\epsilon_f-p)
 \frac{2\Delta}{(p_j-\epsilon_d)^2 + \Delta^2}
\label{chap1RLdotexp}
\ee.

We compare this result to the one from the traditional Bethe-Ansatz,
defined with periodic boundary conditions on a ring of length $L$. In
the usual Bethe-Ansatz, one puts the system on a circle and imposes the
self consistency condition that $|\psi\>$ at $x=0$ equals $|\psi\>$ at
$x_j=L$ \cite{natanlectures}.  This leads to the B.A. equations. For
this model where the two-particle S-matrices are trivial, the
B.A. equations yield for the energy \be E= \sum p_j = \sum_{j} \left(
\frac{2\pi n_j}{L} + \frac{1}{L}\delta_{p_j} \right).  \ee The
$\{n_j\}$ are integers corresponding to the energy of a free electron
and the $\{ \delta_{p_j} \}$ the shift in the energies due to the
impurity.  For future reference, define the 'impurity' energy as
$E_{imp} =\lim_{L\rightarrow \infty} \frac{1}{L} \sum_{j}
\delta_{p_j}= \int dp \rho(p) \delta(p) $ with $\rho(p)$ the
distribution that describes the free electrons in the Bethe-Ansatz
basis. From the Feynman Hellman theorem \cite{Feynman}, we know that
\bea \<n_d\>&=& \frac{\partial E}{\partial \epsilon_d}= \frac{\partial
E_{imp}}{\partial \epsilon_d} = \frac{1}{L}\sum_j
\frac{2\Delta}{(p_j-\epsilon_d)^2 + \Delta^2} \nonumber \\ &=&\int dp
\, \theta (E_f-p) \frac{2\Delta}{(p_j-\epsilon_d)^2 + \Delta^2} \eea
in agreement with the expression we computed using the scattering
state formalism (\ref{chap1RLdotexp}).

It is helpful to define an operator that directly yields the
impurity energy using scattering states. This is done by
considering the overlap of the outgoing scattered waves with the
unscattered Fermi-sea. Define a state $|\Phi^+\>$ that describes
 a bath of
outgoing particles (i.e. all particles are to the right of the
impurity) 
\be |\Phi_o^+ \> = \int \prod_i^n dx_i\prod_{s=1}^N
\theta(x_s) e^{i\sum_{j=1}^{N}p_j x_j} \psi^\dagger(x_1)\ldots
\psi^\dagger(x_N) |0\> \label{RLphi+} \nn \ee 
Then, we can define
impurity energy alternatively in terms of an impurity energy
operator, $\hat{E}_{imp}$, that acts on  scattering state
$|\psi\>$ 
\be 
\hat{E}_{imp}|\Psi\>= \lim_{L \rightarrow \infty}
\frac{-i}{L}\log{\left(\frac{\< \Phi^+|\Psi\>}
{\<\Phi^+|\Phi^+\>}\right)} |\Psi \>. \label{DefEimpOp}\nn 
\ee 
A straightforward calculation shows that the expectation value of
impurity-energy operator 
\be 
\<\hat{E}_{imp} \>= \lim_{L
\rightarrow \infty} \frac{\<\Psi|\hat{E}_{imp}|\Psi
\>}{\<\Psi|\Psi\>} =\frac{1}{L} \sum_j \delta_{p_j} = \int dp
\rho(p) \delta(p) 
\ee 
agrees with the expression derived from
traditional methods. The virtue of this operator is that it can be
generalized in a straightforward manner to all integrable
quantum-impurity models. This object is closely related to the
many-body $T$-matrix for the quantum-impurity model.

{\bf RLM at Finite Temperatures: Thermodynamical Properties}
Consider now the finite temperature case.
At finite temperature, $T>0$, the system is no longer described
by  single scattering eigenstate. Instead,
we must consider a density matrix of the form (\ref{chap1scatdensity})
composed of scattering states weighted by the thermal Boltzmann distribution.
Label the set of $N$-particle scattering states
by the energy of the incoming electrons $\{ |\psi, m \> \}$, with $m$
 labelling
all possible sets of energies for the particles $p_1 < p_2 <\ldots <p_N$.
We expect that these scattering
states are a complete basis  for the Hilbert space,
and indeed find that this assumption
 reproduces known thermodynamic results correctly.

We calculate the finite temperature properties of the RLM using
(\ref{rhotimeindepexp}). The dot expectation value is calculated using
dot occupation operator $\hat{n}_d=d^\dagger d$:
\bea
\< \hat{n}_d \> &=& \frac{{\rm Tr}(\rho_s\hat{n}_d)}{{\rm Tr}\rho_s}
=  \frac{{\rm Tr} \sum_{m} e^{-\beta E_m^0} |\psi, m\>\<\psi, m|
  \hat{n}_d  }{{\rm Tr } \sum_{m}e^{-\beta E_m^0}| \psi, m\>\< \psi, m|} \nonumber \\
&=&  \frac{\sum_{m,n} e^{-\beta E_m^0} \< \psi,n|\psi, m\>\<\psi, m|
  \hat{n}_d |\psi, n\> }{\sum_{m^\prime, n}e^{-\beta E_{m^\prime}^0} |\< \psi, n| \psi,
  m^\prime\>|^2 } \nonumber
\eea
The above expressions simplify when we work in the infinite physical $L$ limit since
we can keep only leading order terms in $L$.
Recall, that $m, m^\prime$ and $n$ are shorthand labels
for the ordered set of energies of the $N$ electrons
$p_1 < p_2 <\ldots <p_N$. Thus, if $m \neq n$, there is at least one
electron in each state with different energy. Furthermore,notice that the
overlaps of a single particle eigenstates (\ref{chap1RL1partES})
given by (\ref{chap1RL1partnorm}) are leading order in $L$ only if the
energies of the two single particle eigenstates coincide. Hence,
we conclude from (\ref{chap1RLstate}) that the leading order in $L$
contribution to the overlap of multi-particle eigenstates
comes from states where {\it all} particles
have the same energy, or in other words, when the two states
are identical. Thus, for the infinite $L$ limit, we can set
$m=m^\prime=n$ in the above expressions to get
\bea
\< \hat{n}_d \> &=& \sum_{n} \frac{e^{-\beta E_n^0}}
{ \sum_{m^\prime}e^{-\beta E_{m^\prime}^0} }
\frac{\<\psi, n| \hat{n}_d |\psi, n\> }{L^N} \nonumber \\
&\equiv& \sum_n P(n) \frac{\<\psi, n| \hat{n}_d |\psi, n\> }{L^N} \nonumber
\\
&=&  \sum_n \frac{P(n)}{L} \sum_{p_j \in \{p_j\}_n}
\frac{2\Delta}{\Delta^2 + (p_j-\epsilon_d)^2} \nonumber \\
&\equiv& \sum_n \frac{P(n)}{L} \sum_{p_j \in \{p_j\}_n}
n_d(\epsilon) \eea where $P(n)$ is the Boltzman probability for
the state labelled by $n$. We can now use a standard trick of
statistical mechanics and replace the sum over all configurations
by an integral over the average occupancy of a level of energy
$p$, $N(p)$, which in this case is given by the finite-temperature
Fermi-Dirac distribution function, $f(p,T)$. This yields \be
\<\hat{n}_d\>= \int dp \, f(p ,T)\frac{2\Delta}{(p_j-\epsilon_d)^2
+ \Delta^2}. \ee Thus, the effect of temperature is then
incorporated by requiring that the momentum distribution of the
incoming electrons be chosen  according to the finite temperature
Fermi-Dirac distribution for free electrons. This expression is in
agreement with known results. We will see that the above argument
is quite general and that the effect of temperature can be
generically incorporated by integrating over finite-temperature
distribution functions instead of their zero-temperature
counterparts.

An almost analogous calculation using the impurity energy operator
$\hat{E}_{imp}$ yields that the finite energy impurity energy
is
\be
\<\hat{E}_{imp} \> = \int dp \, f(p ,T) \delta_p.
\ee
The great limitation of the scattering formalism is that though we can
calculate the finite temperature energy, calculating the free energy is
much trickier. A free energy operator can also be defined for these models
though this is much trickier and will not be discussed in this paper
\cite{unpublished}.


\section{Scattering approach to the Interacting Resonance model (IRLM) Thermodynamics}

In this section, we compute the zero temperature thermodynamic
properties of the interacting Resonance Level Model  (IRLM) within
the scattering framework.\ The IRLM Hamiltonian, \bea H_{\rm
IRLM}&=&H_0 + H_{I}= -i\int dx \psi^\dagger(x) \partial_x \psi(x)
+ H_I\nonumber \\ &=&-i\int dx \, \psi^\dagger(x) \partial_x
\psi(x)+ t(\psi^\dagger(0)d
+ h.c.) \nn \\
&& + U \, \psi^\dagger(0)\psi(0)d^\dagger d + \epsilon_d d^\dagger d \nn
\label{chap1IRLM}
\eea
describes a local level, $d^\dagger$, onto which spinless electrons hop
on and off.  There is an additional Coulomb interaction
between the level and electrons. We consider only the case where
$\epsilon_d >0$, where the level is above the Fermi energy of the electrons.
Unlike the RLM considered earlier, this
model is no longer quadratic and we must use the full Scattering Bethe-Ansatz
(SBA) technology to construct scattering states.

We construct the scattering states. They satisfy the Lippman-Schwinger
equation (\ref{chap1LSE}), and specifying the boundary condition on the
incoming particles, $|\Phi\>$, and the Hamiltonian (\ref{chap1IRLM}),
uniquely determine the corresponding scattering state $|\Psi \>$. In
this section, we restrict ourselves to scattering states where the
incoming particles are a Fermi-sea  at zero
temperature $|\Phi_o\>$.  Such scattering states are sufficient to
describe the zero temperature thermodynamic properties of the IRLM
such as the dot occupation and impurity energy.

In principle, the scattering formalism can also be
used to describe quasi-particle $S$ and $T$
matrices. We defer these topics to future publications as they require
treating more complicated boundary condition
for incoming particles that includes quasi-particle excitations
above the Fermi-sea.

\subsection{Construction of the scattering state}

The scattering states for the IRLM are constructed using the SBA,
directly in open systems of infinite size, $L \rightarrow \infty$.
The most general N-particle eigenstate is of the Bethe-Ansatz form
\bea \hspace*{-.2in}|\{ p \} \> =  A\int d\vec{x}
e^{\frac{i}{2}\sum_{i<j}sgn(x_i-x_j)\Phi(p_i,p_j)} \prod_{j=1}^N
\alpha_{p_j}^\dagger (x_j) |0\> \label{psiIRLM} \eea with \be
\Phi(p.k) = \tan^{-1}{\left(\frac{U(p-k)}{2(p+k-2\epsilon_d)}
\right)} \ee and $\delta_p$ and $e_p$ given  in
(\ref{chap1defphaseshift}) and (\ref{chap1RL1partES}) \cite{IRLM}.
Note that $\alpha^\dagger$ is the operator that creates a
single-particle eigenstate (\ref{chap1defalpha}) in the
non-interacting RLM. The  states $|\{p\} \> $ are a complete set
of states in terms of which a particular scattering state can be
constructed as a linear combination of by  the set $(\{p\})$
determined by the boundary conditions. In our case the boundary
condition requires that the incoming particles look like a free
Fermi sea. As discussed previously, for this boundary condition a
single state $|\{ p \} \> $ with  appropriately chosen set $\{p\}$
suffices to determine $|\Psi \>_s$. In more detail,  when all the
particles are to the left on the impurity, $\{x_i \} <0$, $|\Psi
\>_s$ must reduce to an eigenstate of $H_0$, $|\Phi_o \>$,
describing a zero temperature Fermi sea and a decoupled impurity.
When all the $\{ x_j \} <0$, the operators $ \{
\alpha_{p_j}^\dagger(x_j) \}$ reduce to $\{ e^{ip_j
x_j}\psi^\dagger(x_j) \}$ and the eigenstate $|\Psi \>_s$ reduces
to \bea |\psi \> \rightarrow  \int d\vec{x} A
e^{\frac{i}{2}\sum_{i<j}sgn(x_i-x_j)\Phi(p_i,p_j)} e^{i\sum_{j}
x_j p_j} \prod_{j=1}^N \psi^\dagger(x_j) |0 \>. \label{psix<0} \nn
\eea Thus, we must choose the $\{ p_j \}$ in such a manner that
the above expression describes a free Fermi sea.

Despite its appearance the expression on the right hand side
is an eigenstate of  $H_o$. This can be seen by applying  $h_o= -i
\sum_{j=1}^N \partial_{x_j}$
 to the wave function. Indeed, since all particles are right mover
the scattering S-matric $S= e^{i\Phi(p_i,p_j)}$ describes the choice
of  a Bethe basis in the infinitely  degenerate energy subspace of
free electrons. Thus, for   $\{ x_j \} <0$, $ |\Psi \>_s$ reduces to
 an eigenstate expressed in the Bethe  basis
characterized by the two-particle S-matrix $S=
 e^{i\Phi(p_i,p_j)}$. This Bethe basis is the
 natural basis for our problem since, as discussed previously, degenerate
 perturbation theory  demands that we choose the
 basis for the free electron eigenstates by ``turning off'' the
perturbation, in this case the coupling to the quantum impurity.
It is
worth emphasizing that the momenta $\{p_j \}$ should coincide with the usual Fock
momenta of quasi-particles only when $U=0$ and
the $S^{ij}=1$.

The boundary-condition on
incoming particles must be implemented in the
Bethe-Ansatz basis with a non-trivial two particle electron
S-matrix  $S= e^{i\Phi(p_i,p_j)}$.  As discussed previously,
the requirement that the incoming particles be
a Fermi sea translates in this Bethe basis into the condition that in $| \Phi_o \>$ the
incoming particles be an eigenstate of $H_0$ of the form
(\ref{NpartBAwf})
\begin{align}
|\Phi\>_{bath}
= A &\int d\vec{x} e^{i\sum_j p_jx_j}
e^{\frac{i}{2}\sum_{i<j}sgn(x_i-x_j)\Phi(p_i,p_j)} \nn \\
&\psi^\dagger(x_1)\ldots \psi^\dagger(x_N)|0\>
\label{chap1IRLMPhi}
\end{align}
with the additional condition that the distribution for the BA
momenta of the incoming particles, $\rho(p)$, satisfy a set of
free Bethe-Ansatz equations for an {\em auxiliary}  problem of
free electrons on a ring of length $L^\prime$ with a two particle
S-matrix,
 $S=\exp{(i\Phi(p,k))}$. These equations are derived in the appendix
(\ref{A:freeIRLMBAE}) and are given by,
\bea
\rho(p) &=&\frac{1}{2\pi} -\int dk \rho(k)K(p,k)
 \\
K(p,k)&=& \frac{1}{2\pi}\frac{\partial \Phi(p,k)}{\partial p}
= \frac{U}{\pi}
\,\frac{(\epsilon_d-k)}
{(p+k-2\epsilon_d)^2+\frac{U^2}{4}(p-k)^2}.
\label{chap1:IRLMBAE} \nn
\eea

The desired scattering state,  $|\Psi\>_s$, is  given by
(\ref{psiIRLM}) with {\it the additional requirement} that the
distribution of the BA momenta, $\rho(p)$, solves the Bethe-Ansatz
equation above.  The simplicity of the equation follows from the
fact the ground states in the Fock basis and  in the Bethe basis
are unique. This is no longer the case for excited 
excited states. It is also worth emphasizing that
(\ref{chap1:IRLMBAE}) correspond to a free Hamiltonian $H_0$ and
thus differ from the usual Bethe-Ansatz equations for the
IRLM \cite{IRLM} in that they contain no impurity contribution.

\subsection{Zero Temperature Properties}

Having constructed scattering states, we now use them to calculate the
thermodynamic properties of the IRLM.
In particular, we will use
scattering states to calculate the zero-temperature dot occupation $\< \hat{n}_d \> =
\<d^\dagger d \>$ and the impurity energy $E_{imp}$ defined as
using the impurity energy operator (\ref{DefEimpOp}). At zero temperature,
$E_{imp}$ plays the role of the free-energy for all dot thermodynamic
properties. We then show that our results agree with those derived
using traditional  Bethe-Ansatz techniques.

To calculate the impurity dot occupation we use (\ref{chap1timeindepexp})
which yields
\bea
\<\hat{n}_d \> = \frac{\<\Psi | d^\dagger d |\Psi \>}{\<\Psi|\Psi \>}.
\eea
with $|\Psi \>$ as in (\ref{psiIRLM}). As is usual in scattering
theory, we regularize our calculations by placing the system in a box
of size $L$. Since scattering is defined only for open systems,  the
physical system correspond to the infinite $L$ limit and finite $L$
properties are {\it not} well defined. From the
definition of $\alpha^\dagger$ (\ref{chap1defalpha}),
it follows that $d^\dagger d \alpha_{p_s}^\dagger(x_s)|0\>= \delta(x_s)
e_{p_s} d^\dagger|0\>$. Thus,
\begin{align}
 \< \Psi| d^\dagger d | \Psi \>  &=\sum
(-1)^s
 A^2 \int d\vec{y} \, d\vec{x} \, \delta(x_s) \nn \\
& \times \,e^{\frac{i}{2}\sum_{i<j}(sgn(x_i-x_j)-sgn(y_i-y_j))\Phi(p_i,p_j)}\nonumber \\
&\times \,e_{p_s}\prod_{j^\prime, j=1, j \neq s}^N \<0|
\alpha_{j^\prime}(y_{j^\prime})d^\dagger \alpha_j^\dagger (x_j) |0\>
\label{chap1IRLMtemp1}
\end{align}
A very similar calculation yields
\begin{align}
\< \Psi| \Psi \> =&
 A^2 \int d\vec{y} \, d\vec{x} \,
e^{\frac{i}{2}\sum_{i<j}(sgn(x_i-x_j)-sgn(y_i-y_j))\Phi(p_i,p_j)} \nn \\
& \prod_{j^\prime, j=1}^N \<0|
\alpha_{j^\prime}(y_{j^\prime})\alpha_j^\dagger (x_j) |0\>
\label{chap1IRLMtemp2}
\end{align}

To proceed with the calculation we note that from (\ref{chap1defalpha}), one has the relations
\bea
\{ \alpha_j (x_j) , \alpha_s^\dagger(x_s) \} &=&
e^{i(p_s-p_j)}[\theta(-x_s)+
  e^{i(\delta_{p_s}-\delta_{p_j})}\theta(x_s)] \nn \\
  && \times \, \delta(x_s-x_j)+
e_{p_j}e_{p_s}\delta(x_s)\delta(x_j) \nonumber \\
\{d , \alpha_s^\dagger(x_s) \}&=& e_{p_s}\delta(x_s)
\eea
The right hand side of the first equation has two terms: the first
term proportional to $\delta (x_s-x_j)$ comes from the anti-commutation
of the fermionic field $\psi$ while the second comes from $d$. When
calculating (\ref{chap1IRLMtemp1}) and (\ref{chap1IRLMtemp2}) keeping only the
first term is sufficient to get the leading order in $L$ in the dot
occupation since the first term contains only one delta function where
as the second contains two. In explicitly open systems where $L$
in infinite, it is sufficient to treat the
anti-commutation relation as
\begin{align}
\{ \alpha_j (x_j) , \alpha_s^\dagger(x_s) \} &\approx
e^{i(p_s-p_j)} \\
\times&   [\theta(-x_s)
  e^{i(\delta_{p_s}-\delta_{p_j})}\theta(x_s)]\delta(x_s-x_j) \nn
\end{align}
Then, the norm to leading order in $L$ is  given by
\bea
&& \< \Psi| \Psi \>  = A^2 \sum_{\sigma \in S_N} (-1)^{sgn \sigma} \nn \\
&&\int d\vec{y} \, d\vec{x} \,
e^{\frac{i}{2}\sum_{i<j}(sgn(x_i-x_j)(\Phi(p_i,p_j)-\Phi(p_{\sigma(i)},p_{\sigma(j)}))} \nonumber \\
&&
\prod_{s=1}^N
e^{i(p_s-p_{\sigma (s)})x_s}[\theta(-x_s)+
  e^{i(\delta_{p_s}-\delta_{p_{\sigma (s)}})}\theta(x_s)]\delta(y_{\sigma(s)}-x_s)
\label{IRLMtemp10} \nn
\eea
The integral over $y$ is trivial. As explained in the last section,
the leading order in $L$ contribution to
such an integral comes when $e^{i(p_s-p_{\sigma (s)})x_s}=1$ or
precisely when the permutation $\sigma =1$. In this case the integral
is performed trivially and yields
$\<\Psi|\Psi \>= A^2 L^N$. An analogous calculation using (\ref{chap1IRLMtemp1}) yields
to leading order in $L$ that $\<\Psi|d^\dagger d|\Psi\>=
A^2L^{N-1}\sum_{j=1}^N |e_{p_s}|^2$. Combining these two results yields
\bea
\<\hat{n}_d \> = \frac{\<\Psi | d^\dagger d |\Psi \>}{\<\Psi|\Psi
  \>} = \frac{1}{L} \sum_{s=1}^N |e_{p_s}|^2
  = \frac{1}{L} \sum_{s=1}^N \frac{2\Delta}{\Delta^2+ (
  p-\epsilon_d)^2} \nn
\eea
where we have defined the hybridization $\Delta = t^2/2$. In the,
infinite $L$, infinite $N$ limit, we can replace the sum by an
integral over the distribution of BA momenta for the incoming particles, $\rho(p)$
given by (\ref{A:freeIRLMBAE}) to get
\be
\<\hat{n}_d \> =\int dp \, \rho(p) \frac{2\Delta}{\Delta^2+ (
  p-\epsilon_d)^2}
\label{chap1:scatteringdotoccupation}
\ee
We can also compute the impurity-energy using the impurity energy
operator
\be
E_{imp}=\frac{-i}{L}\log{\left(\frac{\< \Phi^+|\Psi\>}{\<\Phi^+|\Phi^+\>}\right)}
\label{chap1:IRLMEimp}
\ee
where $|\Phi^+ \>$ is the eigenstate of the free bath given by
(\ref{chap1:IRLMfreeBAwf}) with the additional
requirement that all particles be to the right of the impurity:
\bea
|\Phi^+ \>
&=& A \int d\vec{x} \prod_{s=1}^N \theta(x_s) e^{i\sum_j p_jx_j}
e^{\frac{i}{2}\sum_{i<j}sgn(x_i-x_j)\Phi(p_i,p_j)} \nn \\
&&\psi^\dagger(x_1)\ldots \psi^\dagger(x_N) |0 \>.
\label{chap1:IRLMfreeBAwf}
\eea
These correspond to outgoing free Fermi-sea of scattered electrons.
In this case,
\bea
\< \Phi^+|\psi\> &=&\int d\vec{y} \, d\vec{x} \,
e^{\frac{i}{2}\sum_{i<j}(sgn(x_i-x_j)-sgn(y_i-y_j))\Phi(p_i,p_j)} \nn \\
&&\times \,  \prod_{j^\prime, j=1}^N
 e^{-ip_{j^\prime}y_{j^\prime}}\theta(y_{j^\prime})
\<0|\psi(y_{j^\prime})\alpha_j^\dagger (x_j) |0\> \nn \\
 &=&A^2 \sum_{\sigma \in S_N} (-1)^{sgn \sigma}\int d\vec{y} \, d\vec{x} \, \nn \\
&&e^{\frac{i}{2}\sum_{i<j}(sgn(x_i-x_j)(\Phi(p_i,p_j)-\Phi(p_{\sigma(i)},p_{\sigma(j)}))}\nonumber \\
&\times&
\prod_{s=1}^N
e^{i(p_s-p_{\sigma (s)})x_s}
  e^{i\delta_{p_s}}\theta(x_s)\delta(y_{\sigma(s)}-x_s)
\label{chap1IRLMtemp3}
\eea
Once again the  integral over $y$ is trivial and
the leading order in $L$ contribution comes from
when the permutation $\sigma =1$
This yields $\< \Phi^+| \psi \>= A^2(L/2)^N e^{i\sum_{s=1}^N \delta_{p_s}}$.
An almost identical calculation to the one used to calculate
$\<\psi|\psi\>$ gives $\<\Phi^+|\Phi^+\>=A^2(L/2)^N$. Combining these
results and substituting in (\ref{chap1:IRLMEimp}) gives
\be
\< \hat{E}_{imp} \>=\frac{-i}{L}\log{\left(\frac{\<
    \Phi^+|\psi\>}{\<\Phi^+|\Phi^+\>}\right)}
= \frac{1}{L}\sum_{s=1}^N \delta_{p_{s}}
\ee
We can once again replace the sum by integrals over $\rho(p)$ to
get
\be
E_{imp}=\int dp \; \rho(p) \delta_{p}.
\ee

These results  can be checked with those arrived at using the traditional
 Bethe-Ansatz (TBA) \cite{IRLM}. The TBA results are almost
identical to those from the SBA except that the distribution $\rho(p)$
must be replaced by TBA distributions $\rho_I(p)$ that include a contribution
from the impurity,
\bea
E_{imp}&=&\int dp \; \rho_I(p) \delta_{p} \nonumber \\
\<\hat{n}_d \> &=& \int dp \, \rho_I(p) \frac{2\Delta}{\Delta^2+ (
  p-\epsilon_d)^2}
\eea
Since, as pointed out in \cite{IRLM}, the distributions
for the TBA, $\rho_{I}(p)$, differs from the distribution from
the SBA, $\rho(p)$, by a term proportional to $N^{-1}$ where $N$ is the number
of particles, in the $L,N \rightarrow \infty$ limit, the SBA and TBA expressions
coincide.

\section{Scattering approach to the Kondo Thermodynamics}

In this section, we discuss how the scattering Bethe-Ansatz could be
used to calculate interesting physical quantities in the Kondo
model. Due to the complexity of the scattering state for the Kondo
model, doing concrete calculations requires the generalization of many
mathematical methods described in the context of spin chains. In
particular, we discuss the tantalizing possibility that many of the
methods of Maillet, Terras, and collaborators \cite{Maillet} can be generalized to
the Kondo model where they may allow exact calculation of as yet
inaccessible interesting physical quantities such as the
impurity T-matrix.
The section starts with a brief discussion of the
scattering state that captures the thermodynamics of the Kondo
problem. In the next subsection, we discuss a possible mapping between the
Kondo problem and an auxiliary 'abelian' problem similar to the IRLM
model. Finally, we discuss how to calculate quantities in this
auxiliary problem and discuss how this formalism may be generalized.
We concentrate only on the zero temperature properties of the Kondo
model. The generalization to finite temperatures
will be presented later.

\subsection{The Scattering State}

The scattering state for the Kondo model is significantly more
complicated than that for the IRLM. These extra complications stem
from the non-abelian nature of the electron two-particle S-matrices in the
Kondo model, $S_e^{ij}= P^{ij}$. This is already evident in the appendix where we
represent the free-Fermi seas in the Kondo Bethe-Ansatz basis.
We focus on constructing scattering states where the incoming
particles are a free Fermi-sea at zero temperature.
Such scattering states, using a conjecture discussed below, allow one to
recover the zero temperature thermodynamics of the Kondo model using scattering states.

It was shown earlier that for scattering states with the asymptotic boundary
conditions that the incoming particles are a Fermi sea,
that the scattering state $|\Psi \>$ can be described by a single
Bethe-Ansatz wavefunction. The most
general Bethe-Ansatz wavefunctions is of the form
\begin{align}
|\Psi \> =\int d \vec{x} \,e^{\sum_{j} p_j x_j} \sum_Q S^Q
A_{\alpha_1 \ldots \alpha_N \alpha_0}\theta(x_Q)
\prod_{j=1}^N \psi_{\alpha_j}^\dagger(x_j) |0\>.
\label{Ikondoscatstate}
\end{align}
with $S^{Q}$ the product of two-particle S-matrices in the Kondo
model, $S^{ij}= P^{ij}$ for electron-electron scattering and
$S^{i0}=\frac{1+iJP^{i0}}{1+iJ}$ for electron impurity scattering
with
$P^{ij}$ is the permutation matrix that exchanges the spins of
particles $i$ and $j$ \cite{natanlectures}.

The asymptotic boundary conditions that the incoming particles be
a filled Fermi-sea now reduce to choosing the Bethe-Ansatz momenta
$\{ p_j \}$ and the amplitude $A_{\alpha_1 \ldots \alpha_N
\alpha_0}$ so that when all the particles are to the left of the
impurity are scattering state reduces to eigenstate of $H_0$ in
the Kondo Bethe-Ansatz basis describing a filled Fermi-sea. This
state, $|\Phi\>_{baths}$  is extensively discussed in the appendix
and is described by a wavefunction of the form
(\ref{A:FreeKondowf}) with $A_{b_1\ldots b_N}$ given by
(\ref{A:KondoA})and BA momenta $\{ p_j \}$ of the form $\frac{2\pi
n_j}{L}$ with $n_j$ integers running from $-N$ to $0$. The
amplitude is written in terms of solutions to
(\ref{A:FreeKondoBAE}), the spin rapidities $\{\Lambda_\gamma \}$.

When all particles are to the left of the impurity, the scattering
state (\ref{Ikondoscatstate}) reduces to
\begin{align}
&|\Psi \> \rightarrow \\
&\int  d \vec{x} \,e^{\sum_{j} i p_j x_j}
\sum_{Q^\prime} S^{Q^\prime}
A_{\alpha_1 \ldots \alpha_N \alpha_0}\theta(x_{Q^\prime};x_0)
 \prod_{j=1}^N \psi_{\alpha_j}^\dagger(x_j) |0\>. \nn
\label{12Kondobcstate}
\end{align}
with $Q^\prime$ a permutation of the $N^e$ electrons,
$\theta(x_{Q^\prime};x_0)= \theta(x_{Q^\prime(1)} <
\theta_{Q^\prime(2)} < \ldots < x_{Q^\prime(N^e)} < x_o)$ with
$x_0$ the position of the impurity. Since reaching the regions
$Q^\prime$ involves no electron-impurity scattering, the $S^Q$ is
a product of the electron-electron scattering matrix $S^{ij}= P^{ij}$ only.
If we choose the momenta $\{p _j \}$ and amplitude $A_{b_1\ldots b_N}$
as in the paragraph above, (\ref{12Kondobcstate}) reduces to the
desired eigenstate of $H_0$ (\ref{A:FreeKondowf}). Thus, the imposition
of the boundary-conditions follows directly from the representation
of the filled Fermi-sea in the Kondo Bethe-Ansatz basis.

Summarizing,  the full
scattering state is  described by (\ref{12Kondobcstate}) with
the additional conditions that  $A_{b_1\ldots b_N}$
be of the form (\ref{A:KondoA}) with the $\{\Lambda_\gamma \}$
solutions to (\ref{A:FreeKondoBAE}) whose density is given by
(\ref{A:sigmao}) and BA momenta $\{ p_j \}$ of the form $\frac{2\pi
n_j}{L}$ with $n_j$ integers running from $-N$ to $0$.
Choosing the amplitude and BA momenta in this way ensure the
scattering state $|\Psi \>$ reduces to a state describing a filled
Fermi sea $|\Phi \>_{baths}$ for incoming particles.

\subsection{Can we map the Kondo to an abelian quantum-impurity problem?}

Having constructed the scattering state, the next task is to compute
quantum-impurity properties using this state. This task is
significantly more difficult than in the IRLM since the amplitude
$A_{b_1\ldots b_N}$ is written in terms of lowering $B$ operators of
the quantum-inverse scattering method. These operators do
non commute but instead satisfy a complicated algebra. This
makes it difficult to manipulate them \cite{natanlectures}.
For this reason, it is quite desirable to explore the intriguing
possibility that the Kondo problem is in fact equivalent to an auxiliary
quantum-impurity problem similar to the IRLM. The central difference
between the Kondo scattering state and the IRLM is that the scattering
states for the Kondo problem are constructed using non-abelian two
particle S-matrices where as the two-particle S-matrix for the IRLM is
 an abelian phase. We  call models with abelian two-particle
S-matrices, abelian quantum impurity problems.  In this section, we
conjecture that the Kondo problem can indeed be mapped to a very
particular 'abelian' quantum-impurity problem. This abelian
quantum-impurity problem correctly reproduces the thermodynamics of
the Kondo model.  We conjecture that arguments similar to those given by Maillet et al
will show that the abelianization  of the problem extends to all quantities  allowing an easy computation of the scattering properties.

The starting point for the conjecture are the Bethe-Ansatz equations
for the Kondo model. These Bethe-Ansatz equations are derived using
the TBA by considering a quantum impurity on a finite ring  of length
$L$ and imposing
periodic boundary conditions. They are given by
\bea
e^{ip_jL} &= \prod_{\gamma=1}^M \frac{\Lambda_\gamma-1
  +ic/2}{\Lambda_\gamma-1 +ic/2} \label{KondoBAE1} \\
 \prod_{\delta=1, \delta \neq \gamma}
^M \frac{\Lambda_\delta-\Lambda_\gamma
  +ic}{\Lambda_\delta-\Lambda_\gamma
  -ic} &= \left(\frac{\Lambda_\gamma-1
  -ic/2}{\Lambda_\gamma-1 +ic/2}\right)^{N^e}
\left(\frac{\Lambda_\gamma
  -ic/2}{\Lambda_\gamma+ic/2}\right)^{N^i}  \nn
\eea
with the additional information that the energy of the Bethe-Ansatz
wavefunction is  $E= \sum_{j} p_j$. The $\Lambda$ are known as the
spin rapidity and parameterize the $M$ spin-down particles.
We also need the log of these equations which yields
\bea
p_j &=& \frac{2\pi}{L} n_j +  \frac{1}{L} \sum_{\gamma=1}^M \left[
\theta_2(\Lambda_\gamma-1)-\pi \right]  \label{KondoBAE10} \\
N^e \theta_2(\Lambda_\gamma-1) &+& N^{i} \theta_2(\Lambda_{\gamma})
=  -2\pi I_{\gamma} + \sum_{\delta=1}^M
\theta_1(\Lambda_\gamma-\Lambda_\delta) \nn
\eea
with $\theta_n(x) = -2\tan^{-1}{nx/c}$ and $n_j$ and $I_j$ integers
coming from the logarithm and are the charge and spin quantum
numbers respectively. The energy of the eigenstate is  given
\bea
E&=& \sum_j p_j = \sum_j \frac{2\pi}{L} n_j +
\frac{N^e}{L} \sum_{\gamma=1}^M \left[\theta_2(\Lambda_\gamma-1)-\pi
  \right] \nonumber \\
&&\hspace*{-.4in}=\sum_j \frac{2\pi}{L} n_j +
\frac{1}{L}\sum_{\gamma=1}^M \left[
  -2\pi I_{\gamma}-N^i\theta_2(\Lambda_\gamma) +
  \sum_{\delta=1}^M \theta(\Lambda_\gamma-\Lambda_\delta)\right] \nonumber \\
&&\hspace*{-.4in}=\sum_j \frac{2\pi}{L} n_j + \sum_{\gamma=1}^M
  -\frac{2\pi}{L}I_{\gamma}+\frac{N^i}{L}
\sum_{\gamma=1}^{M}-\theta_2(\Lambda_\gamma)
\eea

The first two terms are  the energy of a free-electron gas in the
spin-charge decoupled Kondo basis and the last term is the shift
in the ground state energy due to the Kondo impurity. Previously, we
have defined this as the impurity energy $E_{imp}$. Thus, for the Kondo
problem we can write (suggestively)
\be
E_{imp} =\frac{1}{L} \sum_{j=1}^M \delta_{K}(\Lambda_{\gamma})
\ee
with $\delta_K(\Lambda) = -\theta_2(\Lambda_\gamma)= 2\tan^{-1}{(2
  \Lambda/c)}$.
We can also define a phase $\Phi_{K}(\Lambda_\gamma-\Lambda_\delta)=
\theta_1(\Lambda_\gamma-\Lambda_\delta)$ and  a function
$\tilde{k}(\Lambda)=  D\theta_2(\Lambda-1)$ with $D=N^e/L$. Then, the second
Bethe-Ansatz equation in(\ref{KondoBAE10}) can be derived from the
equation
\be
e^{i\tilde{k}(\Lambda_\gamma)L}=
e^{i\delta_K(\Lambda_\gamma)} \prod_{\delta=1}^M
e^{i\Phi_K(\Lambda_\gamma-\Lambda_\delta)} \nn
\ee
by taking the natural logarithm of both sides. This suggestive notation
is illuminating because the above equation is  of the form
of the self-consistency monodromy equation in
the TBA that leads to the BAE
\be
e^{i\tilde{k}(\Lambda)L}A= Z_j A = \left( S^{jj-1} \ldots S^{j1}
\ldots S^{jM} S^{j0}\ldots S^{jj+1}\right)A \nn
\ee
with $S^{js}=e^{i\Phi_K(\Lambda_j-\Lambda_s)}$ ($j,s \neq 0$)
and $S^{j0}=e^{i\delta_K(\Lambda_\gamma)}$.

  Thus, viewing the
$\Lambda's$ as a function of the $\tilde{k}'s$, we see that the BAE for
the Kondo problem could be derived from another abelian
quantum impurity problem of $M$ electrons with an electron-electron
scattering matrix given by $S^{js}=e^{i\Phi_K(\Lambda_j-\Lambda_s)}$
and electron-impurity scattering matrix given by
$S^{j0}=e^{i\delta_K(\Lambda_\gamma)}$.

The SBA can be applied to this auxiliary quantum-impurity problem in the abelian formulation. The
scattering states are analogous to those of the IRLM model with
$\{\Phi, \delta\} \rightarrow \{\Phi_K, \delta_K \}$.
A straight-forward construction and calculation using the SBA
for this abelian problem yield the correct Kondo thermodynamic
properties. This opens up the possibility that scattering
properties of the Kondo model can be alternatively calculated in this
abelian quantum-impurity model where manipulations of the scattering
states are much easier. The scattering states constructed in the last
section are unwieldily because they are defined in terms of complicated
algebras found in the ABA.

The open problem in this conjecture is how to map operators in the
original Kondo problem to this new abelian quantum-impurity
problems. Such a mapping has been worked out for the Heisenburg
spin-chain by Terras and collaborators \cite{Maillet}.
Due to the close analogy of
the Bethe-Ansatz equations for the Heisenburg spin chain equations,
we expect that a similar mapping of operators can be performed for the
Kondo model. If such a mapping can be fully flushed out, the SBA
should lead to exact solutions for many impurity properties such as
the impurity T and S-matrices.

\section{Conclusions}

This paper outlines a scattering framework for quantum-impurity
models. Generally, constructing scattering states for interacting
impurity models is quite difficult. However, if the model is
integrable, these states can be constructed using the Scattering
Bethe Ansatz developed in this paper. The SBA correctly reproduces
the zero temperature thermodynamic properties of both the Kondo
model and the IRLM. In addition, it raises the exciting
possibility that the Kondo model may be equivalent to an abelian
quantum-impurity problem.

The scattering framework also gives us insight into how the
Bethe-Ansatz works. The impurity physics in any
Bethe-Ansatz basis, always looks like single-particle impurity
phase shifts, $\delta$.  This suggests that the Bethe-Ansatz basis
diagonalizes the lead electrons so that the impurity $T$-matrix
is  a phase shift. The complexity of the problem is shifted from
the impurity-electron interaction to finding an appropriate basis for
free electrons. This observation is essential when using the SBA to calculate
nonequilibrium properties of the Kondo model. We feel
that this new perspective on the Bethe-Ansatz may lead to new physical
insights and is worth exploring in greater detail.

\acknowledgements

We would like to thank  Edouard Boulat, Achim Rosch and especially Avi
Schiller, Sung-Po Chao and Eran Lebanon for many illuminating discussions.

\appendix

\section{ The Bethe-Ansatz Basis for Kondo and IRLM} In general constructing scattering
eigenstates is a formidable task. However, for integrable
quantum-impurity models such as the IRLM and
Kondo Models, scattering states can be constructed using a
generalization of the Algebraic Bethe-Ansatz and quantum Inverse
scattering methods. Consequently, the natural basis for these
scattering states  is not the Fock basis, but rather a new
'Bethe-Ansatz' basis. Central to constructing scattering
eigenstates, is the requirement that far away for the impurity,
the incoming electrons look like a free Fermi sea. In this section,
we show how to represent a Fermi-sea in the Bethe-Ansatz
basis appropriate to the IRLM and Kondo models.
For these models, the impurity forces
two-particle S-matrices $S^{ij}$ to be \cite{natanlectures, IRLM}
\bea
S_{IRLM}&=& e^{i\Phi(p_i,p_j)}=
\exp{\left(i\tan^{-1}{\left(
\frac{U(p_i-p_j)}{2(p_i+p_j-2\epsilon_d)}\right)}\right)} \nonumber \\
S_{Kondo}^{ij} &=& P^{ij}
\label{A:twopartSmat}
\eea
In this appendix we show how to represent
free-electrons in the Bethe-Ansatz basis for each of these models.
Denote these two-basis the IRLM basis and the Kondo basis respectively.

\subsection{Free Electrons in the IRLM Bethe-Ansatz Basis}

We first focus on the IRLM. Particles in the IRLM are spinless and
labelled by their B.A. momenta $p_j$. The IRLM Bethe-Ansatz basis
has a electron-electron S-matrices of the form
(\ref{A:twopartSmat}). A Bethe-Ansatz wavefunction in the IRLM
Bethe-Anatz basis is given by (up to an overall multiplicative
phase independent of the $\{x_j \}$) \bea |N , BA\> &=& A \int
d\vec{x} e^{i\sum_j p_jx_j}
e^{\frac{i}{2}\sum_{i<j}sgn(x_i-x_j)\Phi(p_i,p_j)} \nn\\
&&\psi^\dagger(x_1)\ldots \psi^\dagger(x_N)| 0 \>.
\label{A:IRLMfreeBAwf}
\eea
with $sgn(x)$ the sign function which is equal to $\pm 1$ if $x>0/x<0$.
In writing the above expression, we have used the identity that
$(\theta(-x)+e^{i\Phi}\theta(x))=e^{-i\frac{i}{2}\Phi}
e^{\frac{i}{2}\Phi sgn(x)}$. As discussed in the main text, to find
the B.A. momenta $\{p_j\}$ we consider an auxiliary problem of free electrons
living on a finite ring of size $L^\prime$. In the $L^\prime \rightarrow \infty$
the momenta of the physical and auxiliary problem coincide.

We restrict ourselves to the zero temperature case and
when $\epsilon_d$ is greater than the Fermi-level of the electrons. This
is the case considered in \cite{IRLM}.
To derive the Bethe-Ansatz for the BA momenta distribution functions,
we must equate the wavefunction for the auxiliary
problem on a circle when  a particle $j$ is at $x_j=0$ and at
$x_j=L^\prime$. This gives rise to a Bethe-Ansatz condition of the
form
\be
S^{1j}\ldots S^{j-2j}S^{j-1j}A= S^{jN}\ldots S^{jj+1}e^{ip_jL^\prime}
\label{BAper1}
\ee
which implies that
\be
 \left( S^{jj-1} \ldots S^{j1}  S^{jN}\ldots
S^{jj+1}\right) A= e^{-ip_jL^\prime}A.
\label{monodromyIRLM}
\ee
Plugging in the explicit form of the two-particle
S-matrix for the IRLM from (\ref{A:twopartSmat}), this equation
gives rise to an equation for the BA momenta $\{ p_j \}$ of
the form (noting that we can cancel $A$ from both sides
since it is a constant)
\be
e^{ip_j L^\prime}= e^{i \sum_{s=1}^N \Phi(p_j,p_s)}.
\ee
Taking the log
and multiplying by $-i$ one has
\be
p_j = \frac{1}{L^\prime} \sum_{s=1}^N \Phi(p_j,p_s)
+ \frac{2\pi n_j}{L^\prime}
\label{A:freeIRLMBAEdis}
\ee
with $n_j$ an integer.  Notice that the amplitude $A$ has dropped
out of the equation implying that it may be taken to be any constant.
Notice that the 'free' Bethe-Ansatz equations (\ref{A:freeIRLMBAEdis})
for the BA
momenta of $H_0$ in the IRLM basis can  be
obtained form  the Bethe-Ansatz equations for the full
IRLM Hamiltonian (including impurity interactions) \cite{IRLM}
\be
p_j = \frac{1}{2 L^\prime} \sum_{s=1}^N \Phi(p_j,p_s)
+ \frac{2\pi n_j}{L^\prime} + \frac{N^i}{L^\prime}\delta_p,
\ee
by setting the impurity contribution proportional to $N^i$ equal to zero.

As is usual, we will not be concerned with
solving the discrete version of this equation but instead will solve
for the distribution function, $\rho(p)$ describing the density of
solutions to the equations in an interval $(p, p+dp)$. It is
worth emphasizing that such distributions make
sense  only in the limit $L^\prime \rightarrow \infty$. In this limit,
we can replace the sum by an integral to get
\be
p_j = \int dk \rho(k)\Phi(p_j,k)
+ \frac{2\pi n_j}{L^\prime}.
\ee

In the usual way, an equation for the zero temperature density
$\rho(p)$ is obtained by subtracting the equation for
$p_j$  from that for $p_{j+1}$ and expanding in the difference $\Delta
p= p_{j+1}-p_j$ which yields \cite{natanlectures}
\bea
\rho(p) &=&\frac{1}{2\pi} -\int dk \rho(k)K(p,k)
 \label{A:freeIRLMBAE} \\
K(p,k)&=& \frac{1}{2\pi}\frac{\partial \Phi(p,k)}{\partial p}
= \frac{U}{\pi}
\,\frac{(k-\epsilon_d)}
{(p+k-2\epsilon_d)^2+\frac{U^2}{4}(p-k)^2} \nn
\eea
This equations are valid as long as $\epsilon_d$ is greater than
the Fermi energy of the lead electrons.
Though we do not do it here, we could
also find the distribution of the BA momenta at finite
temperatures by considering the free Thermodynamic
Bethe Ansatz (TBA) equations for $H_0$  corresponding to the free zero
temperature BA equation (\ref{A:freeIRLMBAE}) for $H_0$.

Summarizing, in the IRLM basis, there is a non-trivial two particle
S-matrix between free electrons of the form (\ref{A:twopartSmat}). The
presence of this matrix corresponds to working in a Bethe-Ansatz basis
for the IRLM that is distinct from the usual Fock basis.
In this basis, the eigenstates of $H_0$
are of the form
(\ref{A:IRLMfreeBAwf}) with the multi-particle S-matrices $S^Q$ given as
products of two particle S-matrices of the form (\ref{A:twopartSmat}).
For a free Fermi-sea at zero temperature, the distribution for the BA momenta, $\rho(p)$,
is given  by (\ref{A:freeIRLMBAE}) not the Fermi-Dirac distribution functions.

\subsection{Free Electrons in the Kondo Bethe-Ansatz Basis}

We now concentrate of the wavefunction of a free-Fermi sea at zero temperature
in the Kondo basis. In the Kondo basis, free-electrons have a two-particle
S-matrix $S^{ij}= P^{ij}$ where $P^{ij}$ is the permutation
matrix acting on the spins of electrons $i$ and $j$ \cite{natanlectures}. The
Bethe-Ansatz wavefunction for the Kondo Bethe-Ansatz basis is
\bea
|N \>
= \int d\vec{x} e^{i\sum_j p_jx_j}(S^Q)_{a_1 \ldots a_N}^ {b_1 \ldots b_N}
A_{b_1 \ldots b_N}\theta(\vec{x}_Q) \label{A:FreeKondowf} \\
\psi_{a_1}^\dagger(x_1)\ldots \psi_{a_N}^\dagger(x_N)|0\> \nn
\eea
with $S^Q$ an appropriate product of two particle S-matrices $P^{ij}$.
Note that the amplitude in the region $Q=1$, $A_{b_1 \ldots b_N}$
and the choice of BA momenta $\{ p_j \}$
are still unspecified. We will once again have to choose these
appropriately by considering an auxiliary problem defined on a
circle of length  $L^\prime$. In the limit where $L^\prime \rightarrow
\infty$, the expressions from the auxiliary problem coincide with
those for the infinite-size open system. Thus, we can use the
beautiful quantum-inverse scattering technology \cite{natanlectures}.

Once again the BA equations for the auxiliary problem are derived
by equating the wavefunction for the auxiliary
problem on a circle when  a particle $j$ is at $x_j=0$ and at
$x_j=L^\prime$. This gives rise to a Bethe-Ansatz condition of the
form
\begin{align}
&(Z_j)_{a_1 \ldots a_N}^{b_1 \ldots b_N}
A_{b_1 \ldots b_N}=  \\
&\left( S^{jj-1}  .. S^{j1}  S^{jN} ..
S^{jj+1}\right)_{a_1 \ldots  a_N}^{b_1 \ldots b_N} A_{b_1 \ldots
b_N}= e^{-ip_jL^\prime}A_{a_1 \ldots a_N}.\nn
\end{align}
We must choose $ A_{b_1 \ldots
b_N}$ such that it is eigenvector for the equation
$Z_j A= z_j A$ with eigenvalue $z_j= e^{-ip_j L}$.
Note, that in general there are many solutions to this equation. We
will be concerned with a single eigenvector, namely the ground-state.

A general method called the quantum-inverse scattering method has been
developed to solve this problem. Let
$N$ and $M$ denote the total number of particles and the total number
of spin down particles respectively. Let us denote the spin Hilbert
space of particle $j$ by $V_j$.
Let $V^N \otimes \prod_{j=1}^N V_j$ be the $N$-particle spin-space.
The Bethe-Ansatz equations for the ferromagnetic vacuum are
\cite{natanlectures}
\bea
z_j =
\lambda(\alpha_j, \beta_1 \ldots \beta_M)&=& \prod_{\gamma=1}^M
\frac{\Lambda_\gamma+i\frac{c}{2}}{\Lambda_\gamma-i\frac{c}{2}} \nonumber \\
\prod_{\delta=1, \delta \neq \gamma}^M
\frac{\Lambda_\delta-\Lambda_\gamma+ic}{\Lambda_\delta-\Lambda_\gamma-ic}
&=& \left( \frac{\Lambda_\gamma-i\frac{c}{2}}{\Lambda_\gamma +i\frac{c}{2}}\right)^{N}
\label{A:FreeKondoBAE}
\eea

Each set of solutions to the Bethe-Ansatz equations $\{ \lambda_j \}$
corresponds to a different eigenstate of $H_0$.
Different choices of $M$ correspond to eigenstates
of spin $N/2-M$. Since, We are interested in the ground state configuration
with zero spin we restrict ourselves to the sector where $M=\frac{N}{2}$.
Let the solutions of the BA equations with $M=N/2$ for the ground state be given by
$\{\Lambda_j^{gs}\}$. Define $\beta_\gamma= \Lambda_\gamma^{gs} +ic$  and denote the
ferromagnetic vacuum in the space $V^N$ by $|\omega \> =
\prod_{j=1}^N  \left( \begin{tabular}c
1 \\
0 \end{tabular} \right)_j$.
Then, the amplitude $A_{b_1 \ldots b_N}$ in (\ref{A:FreeKondowf}) is
given by
\begin{align}
&A(\Lambda_1 \ldots\Lambda_M)_{b_1 \ldots b_N} = B(\Lambda_1+ic/2)\ldots
B(\Lambda_M+ic/2)|\omega \> \nn \\
&=\sum_{j_1\ldots j_M} A_{j_1 \ldots j_M}\sigma_{j1}^- \ldots
\sigma_{j_M}^- |\omega>
\label{A:KondoA}
\end{align}
where the usual spin amplitude notation $A_{b_1 \ldots b_M}$ is
written as $A_{j_1 \ldots j_M}$ by specifying the position of the $M$
down spins and the operators $B(\Lambda_\gamma+ic/2)$ are defined in the
as is usual in the quantum-inverse scattering matrix \cite{natanlectures}.
The $B$ are best
thought of as generalized lowering operators $\sigma^-$ that
lower the spin of particle $j$. The BA momenta
$\{ p_j \}$ in (\ref{A:FreeKondowf}) on the other hand are trivially of
the from $p_j =\frac{2\pi n_j}{L}$. For the ground state, it runs from
some lower cut-off $K=-\frac{2\pi N}{L}$ to the Fermi energy at zero.

Just as in the IRLM case, we can talk about the density of $\{\lambda_j \}$,
$\sigma_o(\lambda_j)$ rather than the individual $\lambda$ themselves. This approach is
valid in the limit: $N,L \rightarrow \infty$ with $D=N/L$ held fixed.
To get the distribution for the density,
we take the logarithm of both sides of the second
equation in (\ref{A:FreeKondoBAE}) to get
\be
N^e \theta(2\Lambda_\gamma)= -2\pi I_{\gamma}+ \sum_{\delta=1}^{M}
\theta(\Lambda-\Lambda_\gamma)
\ee
with $\theta(x)=-2\tan^{-1}(x/c)$ and $I_{\gamma}$ an integer. Since
we are interested in the $\{\lambda_j\}$ for the groundstate, we set $M=N/2$.
We then consider $\sigma_o (\Lambda)$ describing
the number of solutions in an interval $(\Lambda,
\Lambda+d\Lambda)$. Standard manipulations yield the equation \cite{natanlectures}
\bea
\sigma_o(\Lambda)&=& \frac{2c}{\pi} \left[
\frac{N}{c^2+4\Lambda^2}\right]-\frac{1}{\pi}\int d \Lambda^\prime
\sigma_o(\Lambda^\prime)K(\Lambda-\Lambda^\prime) \nonumber \\
K(\Lambda) &=& \frac{1}{\pi} \frac{c}{c^2+\Lambda^2}.
\eea
This can be easily solved by Fourier transform to yield
the equation for the density of $\{\lambda_j\}$ in the groundstate,
\be
\sigma_o(\Lambda)= \frac{1}{2c}\frac{N}{\cosh{\frac{\pi}{c}\Lambda}}.
\label{A:sigmao}
\ee

To summarize, a Free Fermi sea in the Kondo Bethe-Ansatz basis is
captured by a state of the form (\ref{A:FreeKondowf}) with $A_{b_1
\ldots b_N}$ given by (\ref{A:KondoA}), the $\{\Lambda_\gamma \}$
solutions to (\ref{A:FreeKondoBAE}) whose density is given by
(\ref{A:sigmao}). The BA momenta $\{ p_j \}$ are of the trivial

\bibliographystyle{prsty.bst}
\bibliography{bibliothesis}

\end{document}